\documentclass[prd,floatfix,onecolumn,amsmath,amssymb]{revtex4-1}
\usepackage{graphicx,color,dcolumn,booktabs,bm}
\usepackage{subfigure}
\usepackage{amssymb}
\usepackage{longtable}
\usepackage{indentfirst}
\usepackage{epsfig,gensymb,siunitx}
\usepackage{epstopdf}   
\usepackage{slashed}  
\usepackage{cases}
\usepackage[pdfpages]{xcolor}
\definecolor{maroon}{RGB}{139,25,150}
\usepackage{multirow}
\usepackage{float}
\usepackage{pgf}
\usepackage{physics}
\usepackage[colorlinks, citecolor=blue,anchorcolor=red,menucolor=red, linkcolor=red,filecolor=red,runcolor=red,urlcolor=blue,frenchlinks=red, urlcolor=blue]{hyperref}

\usepackage{tikz}
\usetikzlibrary{decorations.pathmorphing}

\begin{document}

	\preprint{}
	
\title{\color{maroon}{Implications of exclusive photon leptoproduction measurements for the proton charge-radius puzzle }}

	\author{The MMGPDs Collaboration:\\
	Muhammad Goharipour$^{1,2}$}
\email{muhammad.goharipour@ipm.ir}
\thanks{Corresponding author}
	
\author{Anoushiravan Moradi$^{3}$}

\author{K.~Azizi$^{3,4,2}$}

	
\affiliation{
$^{1}$School of Physics, Institute for Research in Fundamental Sciences (IPM), P.O. Box  19395-5531, Tehran, Iran\\
$^{2}$School of Particles and Accelerators, Institute for Research in Fundamental Sciences (IPM), P.O. Box 19395-5746, Tehran, Iran\\
$^{3}$Department of Physics, University of Tehran, North Karegar Avenue, Tehran 14395-547, Iran\\
$^{4}$Department of Physics, Faculty of Engineering and Natural Sciences, Dogus University, Dudullu-\"{U}mraniye, 34775 Istanbul, T\"urkiye }

\date{\today}

\begin{abstract}

In the present study, we extend our previous analysis of the proton electromagnetic form factors (FFs) extracted from exclusive photon leptoproduction (EP) measurements in kinematic regions where the Bethe--Heitler (BH) process dominates the cross section by including all currently available high-precision EP data from CLAS and Hall~A Collaborations. Using the same phenomenological framework, we investigate the consistency among the different datasets, determine the proton electromagnetic FFs within several fitting scenarios, and extract the corresponding charge and magnetic radii.
A significant tension is found between the CLAS 2018 measurements and the remaining EP data, with the low-$|t|$ region playing a dominant role in this tension. We systematically investigate this behavior through $|t|$-cut scans and assess its impact on the extracted FFs and radii. For all analyses, the extracted proton charge radius is smaller than the Particle Data Group average and most determinations based on elastic electron--proton scattering. However, the results are consistent, within uncertainties, with the PRad measurement and muonic hydrogen spectroscopy. In contrast, the magnetic radius is found to be compatible with the current world average. These results demonstrate that BH-dominated EP measurements provide an independent and complementary approach to determine the electromagnetic structure of the proton and offer additional support for the small-radius solution of the proton charge-radius puzzle.

\end{abstract}

\maketitle

\section{Introduction}\label{sec:one} 

Electromagnetic form factors (FFs) provide direct insight into the spatial distributions of charge and magnetization, and hence, are fundamental quantities in studying the internal structure of the nucleon~\cite{Diehl:2013xca,Hashamipour:2022noy,Ramalho:2023hqd,Pate:2024acz,Yao:2024uej,Wang:2024abv,Hernandez-Pinto:2024kwg,Cheng:2024cxk,Lin:2024rak,Kuzmin:2024ozz,Cheng:2025yij,Alexandrou:2025vto,Yu:2025jcs,Williams:2025fiv,Arbabifar:2026tev,Lee:2026zgo}.
In quantum electrodynamics and hadronic physics, the Dirac and Pauli FFs, $F_1(t)$ and $F_2(t)$, where  $t$ is the squared four-momentum transfer, 
parameterize the nucleon's electromagnetic current.
They are usually expressed in terms of the the Sachs FFs, $G_E(t)$ and $G_M(t)$, which are more directly related to the electric and magnetic spatial distributions in the Breit frame. The slopes of these FFs at $t=0$ are of particular importance, because they are related to the nucleon charge and magnetic radii. These electromagnetic radii play a crucial role in our understanding of the nucleon structure and have attracted great interest both experimentally and theoretically, especially in the context of the long-standing proton charge radius puzzle~\cite{Pohl:2010zza,Lorenz:2012tm,Pohl:2013yb,Carlson:2015jba,Karr:2020wgh} (see Ref.~\cite{Goharipour:2025yxm} and references therein to get more information).

Traditionally, electromagnetic FFs have been extracted from elastic electron--proton scattering experiments through Rosenbluth separation and polarization transfer techniques~\cite{Kelly:2004hm,Qattan:2004ht,Perdrisat:2006hj,Crawford:2006rz,Arrington:2007ux,A1:2010nsl,A1:2013fsc,Zhan:2011ji,Mihovilovic:2016rkr,Ye:2017gyb,Xiong:2019umf,Lin:2021umz,Christy:2021snt,Qattan:2024pco,Williams:2025fiv}. While Rosenbluth measurements provided the separate determinations of
of $G_E$ and $G_M$, polarization-based methods have significantly improved the precision through the determination of their ratio. However, there are some discrepancies at higher momentum transfers that highlight the importance of two-photon exchange effects and other higher-order corrections. In parallel, the electromagnetic FFs have also been interesting quantities for various theoretical and phenomenological approaches to QCD, including
holographic QCD~\cite{Xu:2021wwj}, lattice calculations~\cite{Park:2021ypf},
and phenomenological generalized parton distribution (GPD) models~\cite{Goharipour:2024mbk}.

In recent years, exclusive processes such as exclusive photon leptoproduction (EP), which includes deeply virtual Compton scattering~\cite{Ji:1996nm,Collins:1998be,Goeke:2001tz,Belitsky:2001ns,Belitsky:2010jw,Kriesten:2020wcx}, have emerged as powerful tools to probe the nucleon structure. These processes are sensitive not only to GPDs~\cite{Radyushkin:1997ki,Diehl:2003ny,Ji:2004gf,Belitsky:2005qn,Boffi:2007yc,Guidal:2013rya,Diehl:2015uka,Kumericki:2016ehc,Mezrag:2022pqk,Xie:2023xkz,Guo:2023ahv,Lorce:2025aqp,Boer:2025ixc}, which provide a multidimensional description of the nucleon as well as novel information on its various properties~\cite{Burkardt:2002hr,Kaur:2023lun,Cichy:2024afd,Polyakov:2018zvc,Burkert:2018bqq,Goharipour:2025lep,Martinez-Fernandez:2025jvk}, but also to the electromagnetic FFs themselves,
particularly in kinematic regions dominated by the Bethe--Heitler (BH) mechanism~\cite{Moradi:2025pkp}.
The BH process, in which the real photon is emitted from the lepton line, has a well-known and precisely calculable amplitude that depends explicitly on $F_1(t)$ and $F_2(t)$~\cite{Belitsky:2001ns,Adams:2024pxw,Kriesten:2019jep}. This sensitivity opens the possibility of extracting electromagnetic FFs from EP measurements in a manner complementary to traditional elastic scattering experiments as we showed in our previous study~\cite{Moradi:2025pkp}.
Such alternative approaches are especially valuable in kinematic regimes where conventional measurements are limited. In particular, accessing low-$|t|$ regions with high precision is crucial for constraining the slopes of the FFs, and hence, getting new information on the nucleon radii.
Therefore, the study of electromagnetic FFs through EP processes provides not only an independent cross-check of existing results but also provides a framework for comprehensive understanding of the nucleon electromagnetic structure.

In the present study, we extend our previous analysis~\cite{Moradi:2025pkp} by incorporating a significantly larger set of EP measurements. The earlier analysis was based solely on CLAS Collaboration data obtained with a 5.75-GeV electron beam~\cite{CLAS:2015uuo}. While the main idea, which is the extraction of nucleon electromagnetic FFs from EP measurements as a complementary approach to traditional elastic electron--nucleon scattering experiments, remains unchanged, the inclusion of a wider range of the experimental data provides a more globally constrained determination of the FFs. Such an extended analysis may also reveal modifications in the extracted $t$-dependence of the FFs compared with our previous results.

The phenomenological framework employed in the present work is identical to that used in Ref.~\cite{Moradi:2025pkp}. In the following sections, we first briefly introduce the experimental data included in the analysis and summarize the phenomenological framework. We then perform some detailed $\chi^2$ analyses to extract the electromagnetic FFs and compare the obtained results with those of our previous study as well as with other available parametrizations and analyses. Finally, we determine the corresponding  charge and magnetic radii of the proton and discuss their implications for our understanding of nucleon electromagnetic structure, especially in the context of the proton charge radius puzzle.

\section{Data selection}\label{sec:two}

The present analysis is based on a comprehensive collection of exclusive EP cross section measurements from CLAS and Hall A Collaborations at Jefferson Lab. Compared with our previous study~\cite{Moradi:2025pkp}, which relied only on the CLAS measurements obtained with a 5.75-GeV electron beam~\cite{CLAS:2015uuo}, the current work contains additional high-precision EP data which cover a wider kinematic region. The complete set of measurements included in the analysis is summarized in Table~\ref{tab:Data}. 
The inclusion of these additional data allows for a more accurate determination of the proton electromagnetic FFs. It should also be noted that we use the open database provided by Burkert {\it et al.}~\cite{Burkert:2025gzu} to include these datasets in the analysis.

Since the aim of this work is to extract the proton electromagnetic FFs through the BH contribution to the EP cross section and then calculate the corresponding nucleon radii, special care must be taken to restrict the analysis to kinematic regions where the BH mechanism dominates. As discussed in Ref.~\cite{Moradi:2025pkp}, the experimentally measured EP cross section contains contributions from the BH process, the DVCS process, and their interference. Consequently, the measured cross sections cannot be directly separated into these individual components.

To identify the kinematic regions in which the BH approximation provides a reliable description of the data, we employ the same selection procedure developed in Ref.~\cite{Moradi:2025pkp}. For each experimental data point, the relative difference between the pure BH prediction and the full EP cross section is evaluated using the \texttt{Gepard} framework~\cite{Kumericki:2006xx,Kumericki:2007sa,Kumericki:2009uq,Cuic:2023mki},
\begin{equation}
\frac{\Delta \sigma}{\sigma}=
\frac{\left|\sigma_{\rm EP}-\sigma_{\rm BH}\right|}
{\sigma_{\rm EP}}\,.
\label{Eq1}	
\end{equation}
Only those measurements satisfying $ \Delta \sigma / \sigma \le 5\% $
are considered in the present analysis. In our previous analysis, several increasingly restrictive criteria ($ 1\% $, $ 2\% $, $ 3\% $, $ 4\% $, and $ 5\% $) were investigated and found to yield mutually consistent determinations of the electromagnetic FFs within uncertainties. The $ 5\% $ criterion was, therefore, adopted as the optimal compromise between BH purity and statistical precision. This provides the largest number of usable data points while maintaining the validity of the BH-dominated approximation.
It should be noted that, as in our previous study~\cite{Moradi:2025pkp}, an additional uncorrelated systematic uncertainty of 5\% is taken into account for all included data points to incorporate the uncertainties associated with this selection procedure, though our investigations indicate that this source of uncertainty does not significantly affect the final extracted $ F_1(t) $ and $ F_2(t) $.

In the present study, the analyzed datasets consist of the CLAS measurements reported in Refs.~\cite{CLAS:2015uuo,CLAS:2018bgk} and the Hall A measurements reported in Refs.~\cite{JeffersonLabHallA:2015dwe,Defurne:2017paw}. They have been listed in Table~\ref{tab:Data}, together with their kinematic coverage and the corresponding numbers of data points after applying the aforementioned selection procedure. 
The resulting dataset provides the broadest EP database considered so far for the extraction of proton electromagnetic FFs from BH-dominated EP measurements.
The CLAS measurements provide the largest kinematic coverage in both $x_B$ and $Q^2$, while the Hall A data offer high-precision measurements in more localized regions of phase space. These two types of measurements, therefore, play complementary roles in constraining the form factors: the CLAS data largely determine the global behavior of the fit, whereas the Hall A measurements provide important local constraints due to their superior precision. The total number of data points included in the analysis is 1249.
These data span the ranges $ 0.125 \le x_B \le 0.5 $, $ 0.108 \le -t \le 0.909~{\rm GeV}^2 $, and $ 1.11 \le Q^2 \le 3.982~{\rm GeV}^2 $, where $x_B$ and $Q^2$ represent the Bjorken variable and photon virtuality, respectively,
with beam energies ($ E $) ranging from 3.355 to 5.88 GeV. 
The azimuthal angle between the leptonic and hadronic scattering planes, $\phi$, covers the interval $ \pm [7.30, 97.37]  $, where symbol $  \pm $ indicates that data are included for both positive and negative values of $\phi$.
Relative to the dataset employed in Ref.~\cite{Moradi:2025pkp}, the present compilation significantly enlarges both the statistical sample and the kinematic coverage. This allows a more accurate determination of the proton electromagnetic FFs. 
It should also be noted that the number of data points from the CLAS 2018 experiment~\cite{CLAS:2018bgk} included in our final analysis (and hence the total number of data points) is less than 1249. This reduction arises because we apply kinematic cuts to the CLAS 2018 dataset due to observed tensions with other measurements, as discussed in Sec.~\ref{sec:four}.

\begin{table*}[t]
	\centering
	\setlength{\tabcolsep}{5pt}
	\caption{Experimental EP datasets included in the present analysis. The kinematic coverage is given in terms of Bjorken variable $x_B$, momentum transfer $-t$, photon virtuality $Q^2$, azimuthal angle $\phi$, electron beam energy $E$, and the number of fitted data points $N_{\rm pts}$. The symbol $  \pm $ indicates that data are included for both positive and negative values of $\phi$.}
	
	\begin{tabular}{lcccccc}
		\hline\hline
		Experiment &
		$x_B$ &
		$-t$ (GeV$^2$) &
		$Q^2$ (GeV$^2$) &
		$\phi$ (deg) &
		$E$ (GeV) &
		$N_{\rm pts}$ \\
		\hline
		
		CLAS 2015~\cite{CLAS:2015uuo}
		& [0.126, 0.475]
		& [0.110, 0.450]
		& [1.110, 3.770]
		& $ \pm $ [37.40, 97.37] 
		& 5.75
		& 516 \\
		
		Hall A 2015~\cite{JeffersonLabHallA:2015dwe}
		& [0.336, 0.401]
		& [0.171, 0.372]
		& [1.820, 2.375]
		& $ \pm $ [7.50, 37.50] 
		& 5.76
		& 72 \\
		
		Hall A 2017~\cite{Defurne:2017paw}
		& [0.360, 0.360]
		& [0.180, 0.300]
		& [1.490, 1.500]
		& $ \pm $ [15.00, 45.00] 
		& 3.355
		& 8 \\
		
		Hall A 2017~\cite{Defurne:2017paw}
		& [0.356, 0.361]
		& [0.181, 0.364]
		& [1.740, 2.000]
		& $ \pm $ [7.50, 52.50] 
		& 4.455
		& 42 \\
		
		Hall A 2017~\cite{Defurne:2017paw}
		& [0.356, 0.361]
		& [0.182, 0.363]
		& [1.520, 1.990]
		& $ \pm $ [7.50, 37.50] 
		& 5.55
		& 32 \\
		
		CLAS 2018~\cite{CLAS:2018bgk}
		& [0.125, 0.500]
		& [0.108, 0.909]
		& [1.129, 3.982]
		& $ \pm $ [7.30, 97.30] 
		& 5.88
		& 579 \\
		
		\hline
		
		Total
		& [0.125, 0.500]
		& [0.108, 0.909]
		& [1.110, 3.982]
		& $ \pm $ [7.30, 97.37] 
		& [3.355, 5.880]
		& 1249 \\
		
		\hline\hline
	\end{tabular}
	\label{tab:Data}
\end{table*}

\section{Phenomenological framework}\label{sec:three} 

The present analysis is based on the same phenomenological framework developed in Ref.~\cite{Moradi:2025pkp}. Since the formalism has already been discussed in detail there, only the essential ingredients relevant to the present work are briefly summarized below.

The fourfold differential cross section for exclusive photon leptoproduction, $ e+p \rightarrow e'+p'+\gamma,
 $ can be written as~\cite{Belitsky:2001ns,Belitsky:2010jw,Benali:2020vma}
\begin{equation}
	\frac{d^4\sigma}{dQ^{2}\, dx_{B}\, dt\, d\phi}
	= \frac{\alpha_{\text{QED}}^{3} x_{B} y^{2}}{8\pi e^6 Q^{4} \sqrt{1+\epsilon^{2}}}
	\left(
	\left| \mathcal{T}_{\text{BH}} \right|^{2}
	+ \left| \mathcal{T}_{\text{DVCS}} \right|^{2}
	+ \mathcal{I}
	\right)\,,
	\label{Eq2}
\end{equation}
where $\mathcal{T}_{\rm BH}$ and $\mathcal{T}_{\rm DVCS}$ denote the BH and DVCS amplitudes, respectively, and $\mathcal{I}$ represents their interference contribution. Denoting the four-momenta of the incoming electron, virtual photon, initial proton, recoiling proton, and final real photon by $k$, $q$, $p$, $p'$, and $q'$, respectively, the standard kinematic variables are defined as
\begin{equation}
Q^2=-q^2\,,
\qquad
x_B=\frac{Q^2}{2p\cdot q}\,,
\qquad
t=\Delta^2\,,
\label{Eq3}
\end{equation}
with
\begin{equation}
\Delta=p'-p=q-q'\,.
\label{Eq4}
\end{equation}
Furthermore,
\begin{equation}
\epsilon=\frac{2x_BM}{Q}\,,
\qquad
y=\frac{p\cdot q}{p\cdot k}\,,
\label{Eq5}
\end{equation}
where $M$ denotes the proton mass and $y$ corresponds to the fractional energy loss of the lepton in the target rest frame.

According to Eq.~(\ref{Eq2}), the cross section of the EP process contains contributions from the pure BH and DVCS processes as well as the BH--DVCS interference term. One can neglect the DVCS and BH--DVCS interference contributions to a good approximation in the kinematic regions for which the BH contribution is dominant in the cross section and, therefore, the other ones become sufficiently small. 
Under these conditions, the EP measurements become essentially sensitive to the electromagnetic FFs entering the BH amplitude. In our previous study~\cite{Moradi:2025pkp}, we demonstrated that it is possible to find the kinematic regions in $ x_B $ and $ \phi $ for which the BH contribution almost entirely dominates the EP cross section, e.g., at $Q^2 = 1$ Gev$ ^2 $, $t = -0.1$ Gev$ ^2 $, and $ E=6 $ GeV.

Consequently, by selecting data points corresponding to BH-dominated kinematics, as discussed in the previous section, and adopting suitable parametrizations for the electromagnetic FFs, it becomes possible to determine them through a fit to the experimental cross section data. Once the FFs are extracted, the corresponding charge and magnetic radii of the proton, $r_E$ and $r_M$, can be obtained from the slopes of the Sachs electric and magnetic FFs at $t=0$ according to
\begin{align}
\left<r_{E}^2\right> &= \left.  6 \dv{G_E}{t} \right|_{t=0} \,, \nonumber \\
\left<r_{M}^2\right> &= \left.  \frac{6}{\mu_p} \dv{G_M}{t} \right|_{t=0}\,,
\label{Eq6}
\end{align}
where $\mu_p$ denotes the proton magnetic moment. Note also that the Sachs FFs are related to the Dirac and Pauli FFs as $G_E(t) = F_1(t) - \tau F_2(t)  $ and $ G_M(t) = F_1(t) + F_2(t) $, where $\tau = -t/(4M^{2})$.

As in our previous study, we examine both dipole and $P$-pole parametrization forms for the Dirac ($F_{1}$) and Pauli ($F_{2}$) FFs
\begin{equation}
F_i(t) = \left(1 - \frac{t}{a_j}\right)^{-2}\,,
\label{Eq7}
\end{equation}
\begin{equation}
F_i(t) = \left(1 - \frac{t}{a_j}\right)^{-P_j}\,,
\label{Eq8}
\end{equation}
where $i=1,2$, and $j=E,M $ denotes the electric ($j=E$ for $  i=1 $) and magnetic ($j=M$ for $i=2$) parameters, respectively. For the Pauli form factor $ F_2(t) $, the normalization condition $F_2(0)=\kappa$ is imposed by multiplying the above expression by the anomalous magnetic moment of the proton, $\kappa=1.793$. The parameters $a_j$ and $P_j$ are treated as free fit parameters and are determined from the experimental data.
Furthermore, motivated by the findings of Ref.~\cite{Moradi:2025pkp}, which showed that the available EP data provide only limited sensitivity to $ F_2(t) $, we also perform alternative fits in which $F_2(t)$ is fixed to the Kelly parametrization~\cite{Kelly:2004hm}. In this way, only $F_1(t)$ is allowed to vary. These fits enable us to assess the impact of the poorly constrained $F_2(t)$ contribution on the extracted FFs and the corresponding nucleon radii. 

The determination of the free parameters is performed through a standard $\chi^2$ minimization using the Python interface to the CERN \texttt{MINUIT} package~\cite{James:1975dr}, \texttt{iminuit}~\cite{iminuit}. The fit is carried out simultaneously for all datasets listed in Table~\ref{tab:Data}. Parameter uncertainties are obtained from the covariance matrix returned by the minimization procedure using the Hessian method. The corresponding uncertainties on the extracted form factors and derived quantities are evaluated through standard error propagation with $\Delta\chi^2=1$ which corresponds to a 68\% confidence level.

%

\section{Analysis Results}\label{sec:four} 

After introducing the experimental datasets included in the present study and the theoretical and phenomenological framework we adopted, it is time to present the results obtained for the FFs and radii from the $ \chi^2 $ analysis of the data. We begin with the simplest fitting scenario, in which both the Dirac and Pauli FFs, $F_1(t)$ and $F_2(t)$, are described by the dipole parametrization of Eq.~(\ref{Eq7}). In this case, the fit contains only two free parameters, $a_E$ and $a_M$.

A simultaneous fit to all datasets listed in Table~\ref{tab:Data} results in  a value of $\chi^2/\mathrm{d.o.f.}=2.26$ which indicates a relatively poor overall description of the data. To investigate the origin of this relatively large $\chi^2$, we examined the contributions from the individual experiments and found a significant tension between the CLAS 2018 dataset~\cite{CLAS:2018bgk} and the other datasets. In addition to the unsatisfying quality of the fit, the inclusion of the CLAS 2018 data leads to a substantial modification of the extracted Dirac and Pauli FFs, and hence Sachs FFs $G_{E}(t)$ and $G_{M}(t)$, when compared with established parametrizations such as that of Kelly~\cite{Kelly:2004hm} and Ye {\it et al.} (YAHL18)~\cite{Ye:2017gyb}. As a consequence, the extracted proton charge and magnetic radii become $ r_E = 0.716 \pm 0.002~\mathrm{fm}
 $ and $ r_M = 1.182 \pm 0.020~\mathrm{fm} $
which are far from the current determinations from elastic electron--proton scattering~\cite{Goharipour:2025yxm} and the Particle Data Group (PDG)  average~\cite{ParticleDataGroup:2024cfk}.

Motivated by these observations, we perform an additional fit in which the CLAS 2018 data are excluded from the analysis. This allows us to assess the impact of this dataset on the extracted FFs and derived observables. In the following, the two analyses are referred to as the ``CLAS18inc" and the ``CLAS18exc", respectively. Table~\ref{tab:chi2_dipole} summarizes the corresponding values of $\chi^2$ together with the individual contributions from each experimental dataset and the number of fitted data points. When all available data are included (CLAS18inc), the dominant contribution to the total $\chi^2$ originates from the CLAS 2018 measurements, which alone contribute $\chi^2=1935.3$ for 579 data points, corresponding to $\chi^2/N_{\rm pts}=3.34$. After excluding the CLAS 2018 data (CLAS18exc), the overall fit quality improves dramatically, yielding $\chi^2/\mathrm{d.o.f.}=1.02$. Moreover, substantial improvements are observed for several Hall A datasets. For example, the Hall A 2017 measurements at beam energies of 4.455 and 5.550 GeV improve from $\chi^2/N_{\rm pts}=1.99$ and $4.10$ to $0.95$ and $1.48$, respectively. A similar, but lesser, improvement is observed for the Hall A 2015 data. If one calculates the proton charge and magnetic radii using the extracted FFs of fit CLAS18exc, more resalable values are obtained, $ r_E = 0.819 \pm 0.011~\mathrm{fm}
 $ and $ r_M = 0.823 \pm 0.018~\mathrm{fm} $.  

\begin{table*}[!tb]
\centering
\setlength{\tabcolsep}{10pt}
\caption{Contributions of the individual experimental datasets to the total $\chi^2$ for the dipole fit. Results are shown for fits performed with and without the CLAS 2018 dataset~\cite{CLAS:2018bgk}, namely ``CLAS18inc" and ``CLAS18exc", respectively. }
\begin{tabular}{lcccccc}
\hline\hline
&
&
\multicolumn{2}{c}{CLAS18inc}
&
\multicolumn{2}{c}{CLAS18exc}
\\
Experiment & $N_{\rm pts}$ & $\chi^2$ & $\chi^2/N_{\rm pts}$ & $\chi^2$ &
$\chi^2/N_{\rm pts}$ \\
\hline

CLAS 2015~\cite{CLAS:2015uuo} 
& 516 & 464.26 & 0.900 & 434.20 & 0.841  \\

Hall A 2015~\cite{JeffersonLabHallA:2015dwe}
& 72 & 196.07 & 2.723 & 153.20 & 2.128 \\

Hall A 2017 ($E=3.355$ GeV)~\cite{Defurne:2017paw}
& 8  & 4.09 & 0.511 & 6.09 & 0.762 \\

Hall A 2017 ($E=4.455$ GeV)~\cite{Defurne:2017paw}
& 42 & 83.66 & 1.992 & 39.75 & 0.946 \\

Hall A 2017 ($E=5.550$ GeV)~\cite{Defurne:2017paw}
& 32 & 131.16 & 4.099 & 47.26 & 1.477 \\

CLAS 2018~\cite{CLAS:2018bgk}
& 579 & 1935.28 & 3.342 & --- & --- \\

\hline

Total & 1249 & 2814.52 & $\chi^2/\mathrm{d.o.f.}=2.26$ & 680.50 & $\chi^2/\mathrm{d.o.f.}=1.02$ \\

\hline\hline
\end{tabular}
\label{tab:chi2_dipole}
\end{table*}

\begin{figure}[!tb]
    \centering
\includegraphics[scale=0.7]{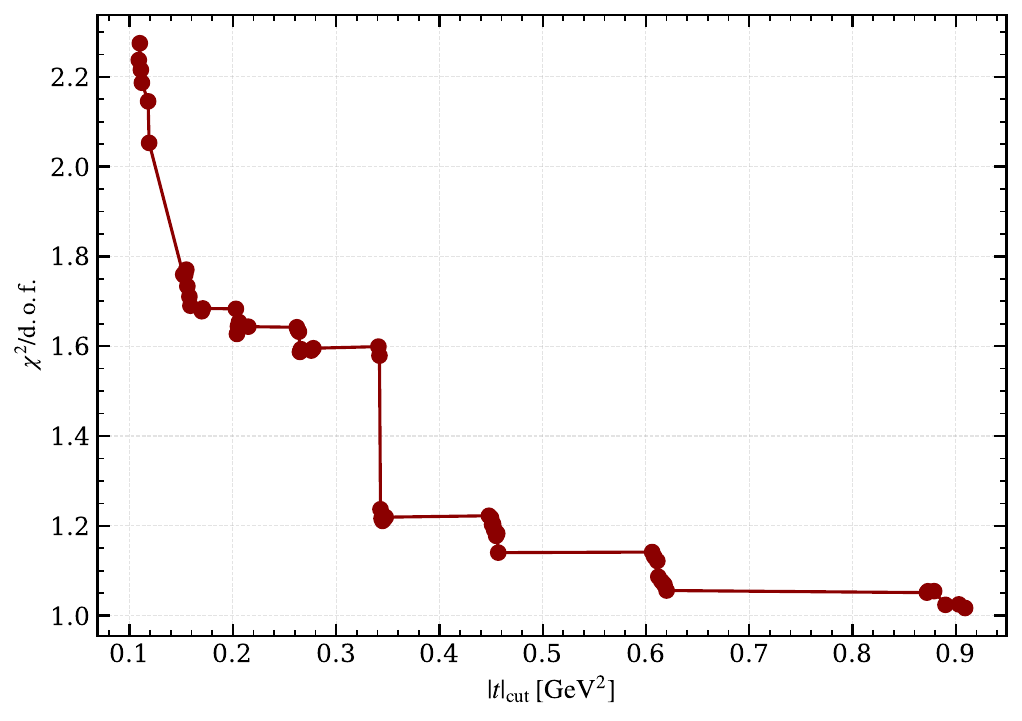}   
\caption{The value of total $\chi^{2}/\mathrm{d.o.f.}$ obtained from the simultaneous dipole fit of the experimental datasets listed in Table~\ref{tab:Data} as a function of the lower momentum-transfer cut, $|t|_{\rm cut}$, applied to the CLAS 2018 dataset~\cite{CLAS:2018bgk}. For each value of $|t|_{\rm cut}$, all CLAS 2018 data points with $|t|<|t|_{\rm cut}$ are excluded from the fit, while the remaining experimental datasets are kept unchanged.  }
\label{fig:tcut_scan}
\end{figure}

To further investigate the tension between the CLAS 2018 data and the remaining EP measurements, we perform a series of additional fits in which parts of the low-$|t|$ CLAS 2018 data are excluded progressively (we exclude data points with the same value of $|t|$ in each step). To be more precise, for a given value of $|t|_{\rm cut}$, all CLAS 2018 data points satisfying $ |t| < |t|_{\rm cut} $ are removed from the fit, while all other datasets remain unchanged. The resulting values of $\chi^2/\mathrm{d.o.f.}$ are shown in Fig.~\ref{fig:tcut_scan} as a function of $|t|_{\rm cut}$.

As can be seen, the fit quality improves steadily as the low-$|t|$ CLAS 2018 measurements are excluded. In particular, $\chi^2/\mathrm{d.o.f.}$ decreases from 2.26 for the fit including the full CLAS 2018 dataset to values approaching unity for sufficiently large values of $|t|_{\rm cut}$. This behavior indicates that the low-$|t|$ region of the CLAS 2018 measurements plays an important role in the observed tension within the present phenomenological framework. Based on the results of the scan, we select two representative values,
$ |t|_{\rm cut}=0.265~{\rm GeV}^2 $ and $ |t|_{\rm cut}=0.343~{\rm GeV}^2 $
which lead to $\chi^2/\mathrm{d.o.f.}=1.59$ and 1.24, respectively, and call the corresponding analyses ``CLAS18inc\_tcut1" and ``CLAS18inc\_tcut2." These choices provide a useful compromise between improving the fit quality and retaining a fraction of the CLAS 2018 data. The number of CLAS 2018 data points retained in the fit is 193 for
$|t|_{\rm cut}=0.265~\mathrm{GeV}^2$ and 142 for
$|t|_{\rm cut}=0.343~\mathrm{GeV}^2$. Table~\ref{tab:dipole_params} summarizes the optimal values obtained for $a_E$ and $a_M$ parameters together with their uncertainties for the four fitting scenarios. The quoted uncertainties correspond to the  $68\%$ confidence level and are determined using the Hessian method. 
\begin{table}[th!]
\centering
\caption{Best-fit values of the dipole parameters $a_E$ and $a_M$ appearing in Eq.~(\ref{Eq7}) obtained from different fitting scenarios, corresponding to different treatments of the CLAS 2018 dataset.}
\label{tab:dipole_params}
\begin{tabular}{c 
        @{\hspace{1.5cm}} c 
        @{\hspace{1.5cm}} c}
\hline\hline
\textbf{Fit} & \textbf{$\boldsymbol{a_E}$ (GeV$^{2}$)} & \textbf{$\boldsymbol{a_M}$ (GeV$^{2}$)} \\
\hline
CLAS18inc & 1.188 $\pm$ 0.010 & 0.239 $\pm$ 0.009 \\[3pt]
CLAS18exc & 0.846 $\pm$ 0.029 & 0.625 $\pm$ 0.048 \\[3pt]
CLAS18inc\_tcut1 & 0.921 $\pm$ 0.042 & 0.569 $\pm$ 0.058 \\[3pt]
CLAS18inc\_tcut2 & 0.851 $\pm$ 0.029 & 0.663 $\pm$ 0.044 \\
\hline\hline
\end{tabular}
\end{table}

\begin{figure}[!tb]
\centering
\includegraphics[scale=0.6]{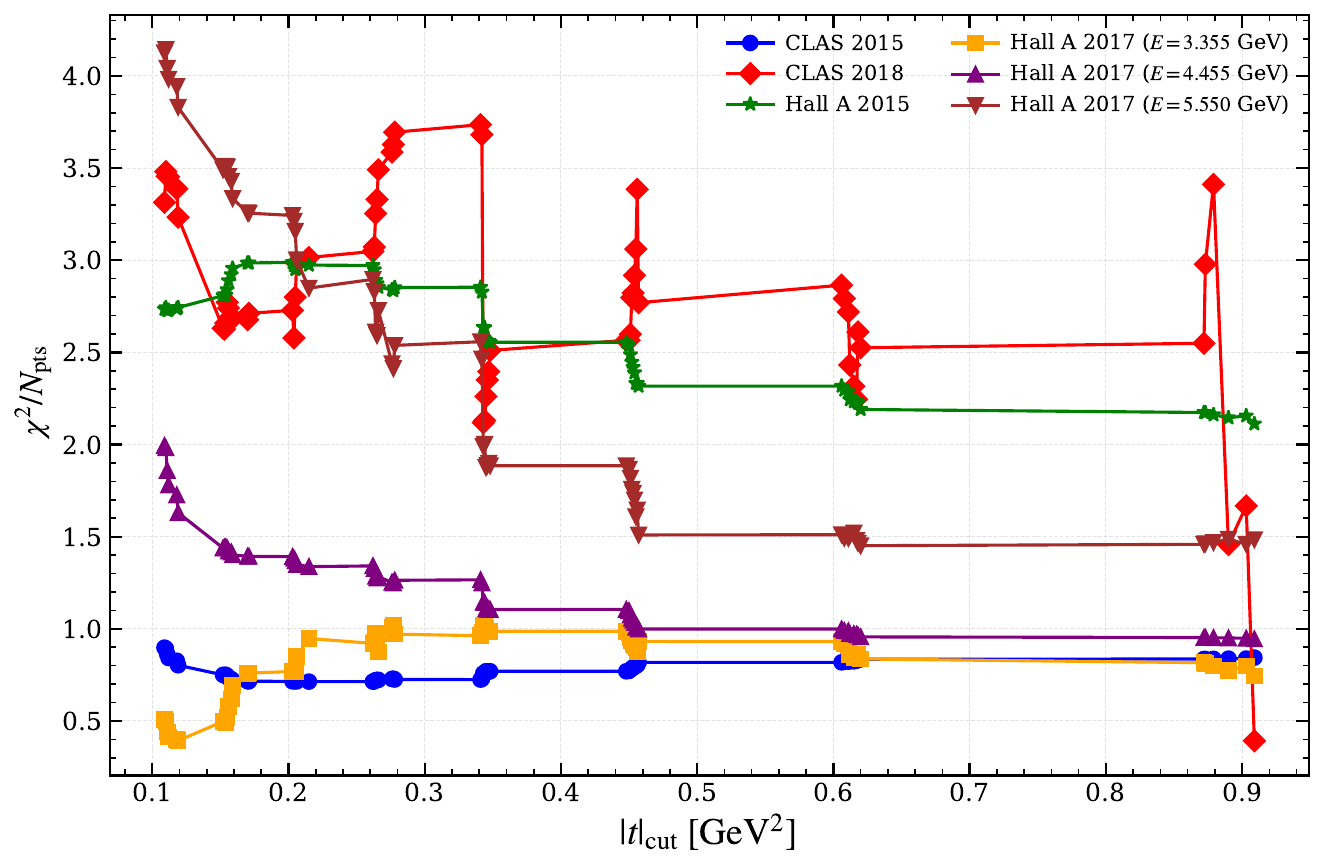}
\caption{The values of $\chi^2/N_{\rm pts}$ for the individual experimental datasets included in the analysis (see Table~\ref{tab:Data}) as functions of $|t|_{\rm cut}$ applied to the CLAS 2018 dataset~\cite{CLAS:2018bgk}. For each value of $|t|_{\rm cut}$, the CLAS 2018 data points satisfying $|t|<|t|_{\rm cut}$ are excluded, while all other datasets are kept unchanged.}
\label{fig:tcut_individual}
\end{figure}

To further clarify the origin of the tension observed between CLAS 2018 and other datasets, we examine how the fit quality of each individual experimental dataset changes as the low-$|t|$ CLAS 2018 points are gradually excluded. The corresponding values of $\chi^2/N_{\rm pts}$ for the individual datasets are shown in Fig.~\ref{fig:tcut_individual}. Note that, in principle, an improvement of the total $\chi^2/\mathrm{d.o.f.}$ could simply be due to removing data points that have large individual contributions to $\chi^2$. Therefore, Fig.~\ref{fig:tcut_individual} provides a more detailed diagnostic of the $|t|_{\rm cut}$ scan presented in Fig.~\ref{fig:tcut_scan}.

As shown in Fig.~\ref{fig:tcut_individual}, the improvement of the global fit is not simply due to reducing the number of CLAS 2018 data points. While the $\chi^2/N_{\rm pts}$ of the remaining CLAS 2018 data changes as the cut is increased, substantial changes are also observed in the $\chi^2/N_{\rm pts}$ values of several independent Hall A measurements. For example, for the full CLAS 2018 dataset, the Hall A 2017 measurements at $E=4.455$ and $5.550$ GeV give $\chi^2/N_{\rm pts}=1.99$ and $4.10$, respectively, according to Table~\ref{tab:chi2_dipole}. When the low-$|t|$ CLAS 2018 data below $|t|_{\rm cut}=0.343~\mathrm{GeV}^2$ are excluded, these values decrease to $1.14$ and $1.99$, respectively. A similar improvement is observed for the Hall A 2015 data, while the CLAS 2015 contribution remains close to unity throughout the scan. Note also that the $\chi^2/N_{\rm pts}$ 
for the CLAS 2018 dataset itself does not have a descending manner, as $|t|_{\rm cut}$ increases and fluctuates around 3. 
Thus, the reduction of the total $\chi^2/\mathrm{d.o.f.}$ cannot be attributed solely to the removal of CLAS 2018 points with large residuals. Rather, the results indicate that the low-$|t|$ CLAS 2018 measurements pull the common fit parameters toward a region that is less compatible with several of the other independent EP measurements.

This observation provides evidence for a dataset tension within the present phenomenological framework, but it does not identify its physical origin.  Actually, the origin of the tension may involve experimental systematics, differences between the datasets, or limitations of the phenomenological framework adopted here. Resolving this issue would require a dedicated study beyond the scope of the present work. For this reason, the fits with the $|t|_{\rm cut}$ values specified above are used primarily as diagnostic  studies. They allow us to quantify the sensitivity of the extracted electromagnetic FFs and proton radii to the observed dataset tension while retaining part of the CLAS 2018 measurements. Therefore, the results obtained from the CLAS 2018-excluded analysis (CLAS18exc), and the two $|t|_{\rm cut}$ scenarios are presented and compared throughout the remainder of this work, rather than treating any one of these scenarios as uniquely preferred by the present analysis.

\begin{figure}[!tb]
    \centering
\includegraphics[scale=0.5]{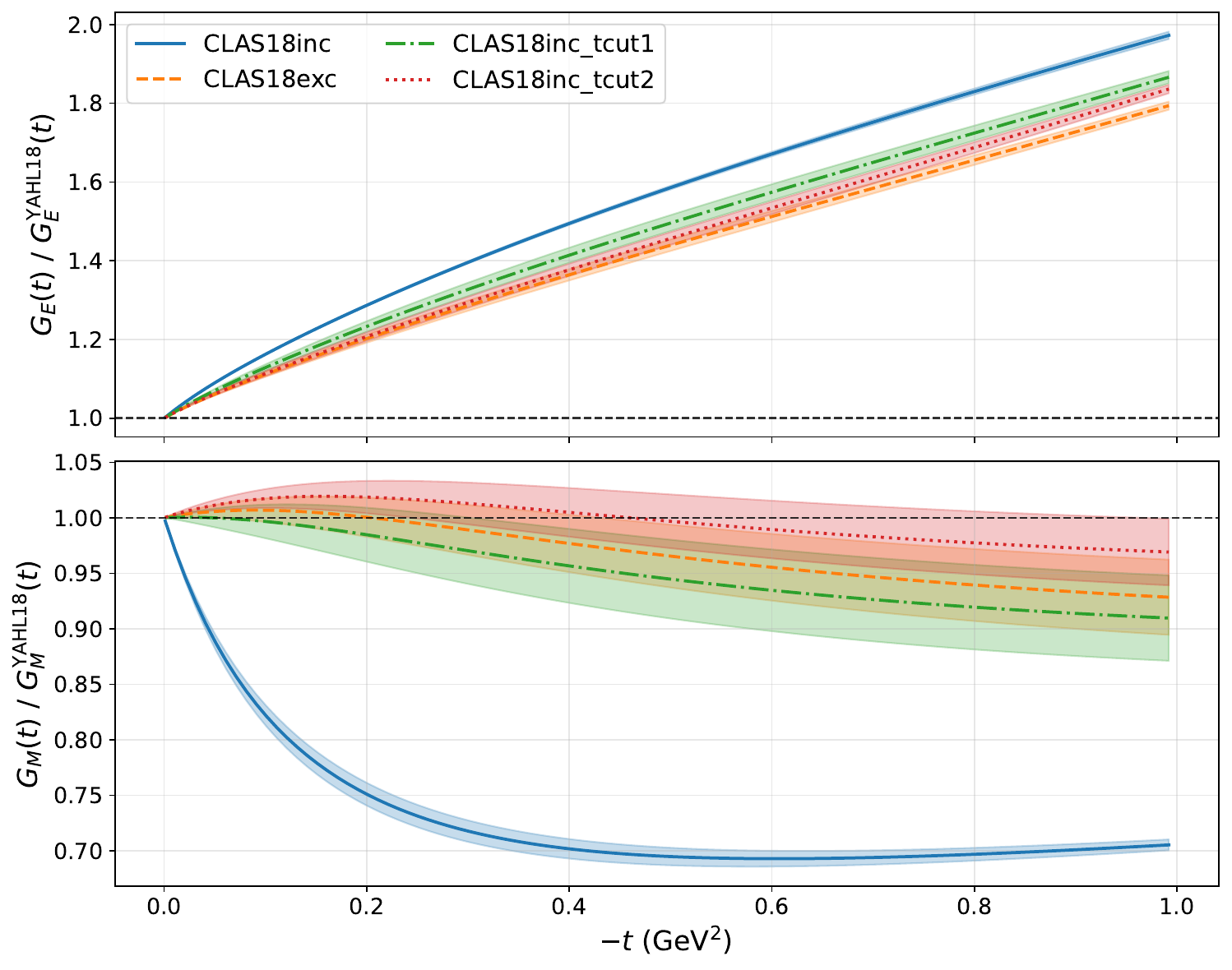}   
\caption{Comparison of the Sachs electric and magnetic FFs, $G_E(t)$ (upper panel) and $G_M(t)$ (lower panel), obtained using the dipole parametrization of Eq.~(\ref{Eq7}) for different fitting scenarios, corresponding to different treatments of the CLAS 2018 dataset, with the YAHL18 parametrization of Ref.~\cite{Ye:2017gyb}. The results are presented as ratios to the YAHL18 parametrization in order to emphasize the differences between the various determinations. }
\label{fig:dipole}
\end{figure}

Figure~\ref{fig:dipole} shows the results for the Sachs FFs $G_E(t)$ and $G_M(t)$ obtained from the four analyses discussed above, compared with the parametrization of YAHL18~\cite{Ye:2017gyb} extracted from elastic electron--proton scattering data. To emphasize the differences, the results are plotted in terms of ratios with respect to the YAHL18 parametrization.
As can be seen, the extracted electric form factor $G_E(t)$ is systematically larger than the corresponding YAHL18 results in all fitting scenarios. The deviation increases with increasing $|t|$ which indicates that the EP measurements are more consistent with milder suppression of $G_E(t)$ 
in the higher momentum-transfer region. The comparison is shown in the range $0 \le |t| \le 1.0~\mathrm{GeV}^2$, which is consistent with the kinematic coverage of the datasets listed in Table~\ref{tab:Data}.

In the case of the magnetic form factor $G_M(t)$, the agreement with YAHL18 is generally better for most fits, with the exception of the CLAS18inc analysis. In this case, noticeable deviations are observed over the full $|t|$ range. The remaining fit scenarios (CLAS18exc and the two $|t|_{\rm cut}$ cases) show only mild differences from YAHL18.
These observations are consistent with the behavior discussed earlier: the inclusion of the full CLAS 2018 dataset without applying a low-$|t|$ cut leads to a significant modification of the extracted FFs, which in turn affects the determination of the proton charge and magnetic radii. Overall, CLAS 2018 data lead to a significant decrease in the uncertainties. This can be observed from Fig.~\ref{fig:dipole} and Table~\ref{tab:dipole_params}. 

We next consider the more flexible $P$-pole parametrization for $F_1(t)$ and $F_2(t)$ defined in Eq.~(\ref{Eq8}). In principle, this parametrization introduces four free parameters, $a_E$, $a_M$, $P_E$, and $P_M$. However, considering the limited sensitivity of the available EP data to the Pauli form factor $F_2(t)$, as discussed in Ref.~\cite{Moradi:2025pkp}, a simultaneous determination of all four parameters is not feasible.
To overcome this limitation, we consider two alternative fitting scenarios. In the first, we impose the constraint $a_E=a_M\equiv a$, while allowing $P_E$ and $P_M$ to vary independently. In the second, we assume $P_E=P_M\equiv P$, leaving $a_E$ and $a_M$ as free parameters. For each scenario, we perform the CLAS18exc, CLAS18inc\_tcut1, and CLAS18inc\_tcut2 analyses and label them with ``$ a $" and ``$ P, $" e.g., CLAS18exc\_$ a $.  The CLAS18inc analysis is not considered, since, as discussed above, the inclusion of the full CLAS 2018 dataset leads to unreliable results.

The resulting values of $\chi^2/\mathrm{d.o.f.}$ for the three fits with the constraint $a_E=a_M$ are 1.01, 1.55, and 1.19, respectively. Identical values are obtained for the fits with the constraint $P_E=P_M$. These values are slightly lower than those obtained with the dipole parametrization (1.02, 1.59, and 1.24, respectively), which reflects the additional flexibility of the $P$-pole form. The corresponding best-fit parameters and their uncertainties are summarized in Table~\ref{tab:fit_params_ppole}. As can be seen, the inclusion of the CLAS 2018 dataset leads to a considerable reduction in the uncertainties of the model parameters in each scenario. However, compared to the dipole case, the model uncertainties are larger due to the increased number of free parameters and the resulting correlations among them.
\begin{table*}[!tb]
    \centering
    \setlength{\tabcolsep}{10pt}
    \caption{Best-fit values of the parameters of the $P$-pole parametrization in Eq.~(\ref{Eq8}) for the two fitting scenarios considered in this work: (i) $a_E=a_M\equiv a$ with free $P_E$ and $P_M$, and (ii) $P_E=P_M\equiv P$ with free $a_E$ and $a_M$. Results are shown for the CLAS18exc, CLAS18inc\_tcut1, and CLAS18inc\_tcut2 analyses.}
    \begin{tabular}{lcccc}
        \hline\hline
        & \multicolumn{2}{c}{Scenario I: $a_E=a_M\equiv a$} 
        & \multicolumn{2}{c}{Scenario II: $P_E=P_M\equiv P$} \\
        \cline{2-5}
        Fit & Parameter & Value & Parameter & Value \\
        \hline
        
        CLAS18exc
        & $a$ & 0.555 $\pm$ 0.056 & $a_E$ & 0.619 $\pm$ 0.062 \\
        & $P_E$ & 1.388 $\pm$ 0.126 & $a_M$ & 0.433 $\pm$ 0.056 \\
        & $P_M$ & 1.841 $\pm$ 0.171 & $P$ & 1.516 $\pm$ 0.126 \\
        
        \hline
        
        CLAS18inc\_tcut1
        & $a$ & 0.591 $\pm$ 0.026 & $a_E$ & 0.716 $\pm$ 0.033 \\
        & $P_E$ & 1.378 $\pm$ 0.054 & $a_M$ & 0.411 $\pm$ 0.030 \\
        & $P_M$ & 2.103 $\pm$ 0.102 & $P$ & 1.603 $\pm$ 0.055 \\
        
        \hline
        
        CLAS18inc\_tcut2
        & $a$ & 0.579 $\pm$ 0.026 & $a_E$ & 0.693 $\pm$ 0.031 \\
        & $P_E$ & 1.373 $\pm$ 0.057 & $a_M$ & 0.413 $\pm$ 0.032 \\
        & $P_M$ & 2.040 $\pm$ 0.097 & $P$ & 1.583 $\pm$ 0.055 \\
        
        \hline\hline
    \end{tabular}
    \label{tab:fit_params_ppole}
\end{table*}

\begin{figure}[!tb]
    \centering
\includegraphics[scale=0.5]{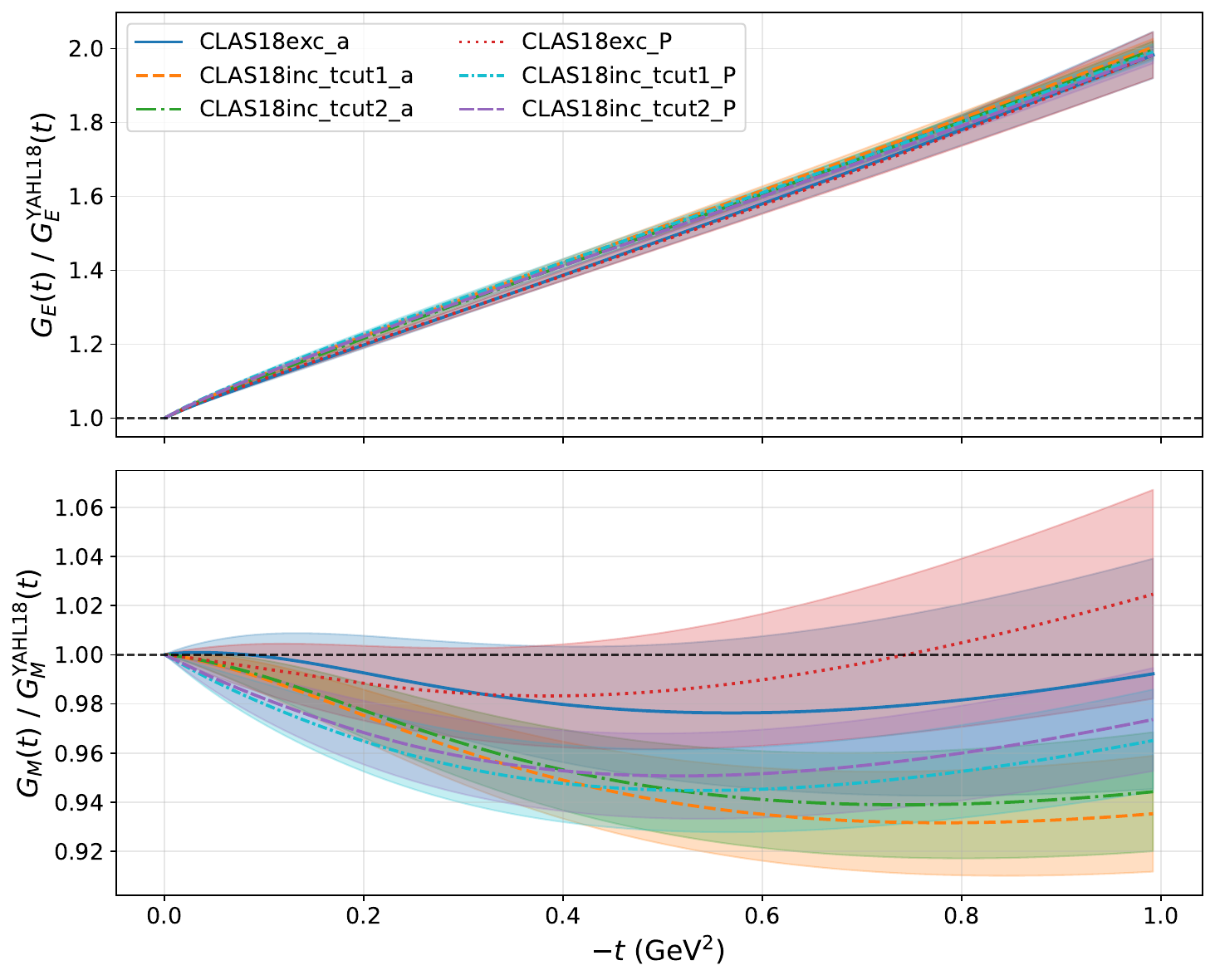}   
\caption{Comparison of the Sachs electric and magnetic form factors, $G_E(t)$ (upper panel) and $G_M(t)$ (lower panel), obtained using the $P$-pole parametrization of Eq.~(\ref{Eq8}) for different fitting scenarios, corresponding to different treatments of the CLAS 2018 dataset, with the YAHL18 parametrization of Ref.~\cite{Ye:2017gyb} extracted from a global analysis of elastic electron--proton scattering data. Three of the fits are performed under the constraint $a_E=a_M\equiv a$, while the remaining three assume $P_E=P_M\equiv P$. }
\label{fig:Pipole}
\end{figure}

Figure~\ref{fig:Pipole} compares the Sachs FFs obtained from the six analyses based on the $P$-pole parametrization. Three of these analyses correspond to the scenario in which the parameter $a$ is fixed ($a_E=a_M\equiv a$), using the CLAS18exc, CLAS18inc\_tcut1 ($|t|_{\rm cut}=0.265~\mathrm{GeV}^2$), and CLAS18inc\_tcut2 ($|t|_{\rm cut}=0.343~\mathrm{GeV}^2$) datasets. The remaining three analyses correspond to the scenario in which the exponent is fixed ($P_E=P_M\equiv P$). As in Fig.~\ref{fig:dipole}, the results are presented as ratios to the YAHL18 parametrization~\cite{Ye:2017gyb} to facilitate the comparison.

The electric form factor $G_E(t)$ exhibits a behavior very similar to that obtained with the dipole parametrization in Fig.~\ref{fig:dipole}. The six analyses are in excellent agreement with each other over the entire $|t|$ range. However, they all predict systematically larger values than the YAHL18 parametrization so that the deviation increases as $|t|$ increases.
The situation is somewhat different for the magnetic form factor $G_M(t)$. Although all six analyses remain in reasonable agreement with the YAHL18 parametrization, with deviations generally less than $7\%$, they are more different from each other and have larger uncertainties than those of $G_E(t)$. This behavior indicates the limited sensitivity of the present EP data to the Pauli form factor $F_2(t)$, and hence poor constraints on $G_M(t)$. Nevertheless, the additional flexibility provided by the $P$-pole parametrization leads to a modest improvement in the quality of the fit, as indicated by the reduced values of $\chi^2/\mathrm{d.o.f.}$.

As discussed above, the EP measurements included in the present analysis cannot provide sufficient sensitivity to constrain the Pauli form factor $F_2(t)$. Therefore, following the strategy adopted in our previous study~\cite{Moradi:2025pkp}, we also perform the CLAS18exc, CLAS18inc\_tcut1 ($|t|_{\rm cut}=0.265~\mathrm{GeV}^2$), and CLAS18inc\_tcut2 ($|t|_{\rm cut}=0.343~\mathrm{GeV}^2$) analyses by fixing $F_2(t)$ to the Kelly parametrization~\cite{Kelly:2004hm} and fitting only $F_1(t)$ using the $P$-pole parametrization. The resulting values of $\chi^{2}/\mathrm{d.o.f.}$ are 1.01, 1.55, and 1.19 for the CLAS18exc, CLAS18inc\_tcut1, and CLAS18inc\_tcut2 analyses, respectively. The corresponding best-fit parameters are
$a_E = 0.580 \pm 0.085$ and $P_E = 1.440 \pm 0.176$,
$a_E = 0.600 \pm 0.039$ and $P_E = 1.450 \pm 0.075$, and
$a_E = 0.580 \pm 0.038$ and $P_E = 1.427 \pm 0.073$.
As can be seen, the values of $\chi^{2}/\mathrm{d.o.f.}$ are identical to those obtained from the $P$-pole fits. However, 
the best-fit values of the parameters and their uncertainties exhibit some modest changes, especially for $ P_E $.

\begin{figure}[!htb]
    \centering
\includegraphics[scale=0.53]{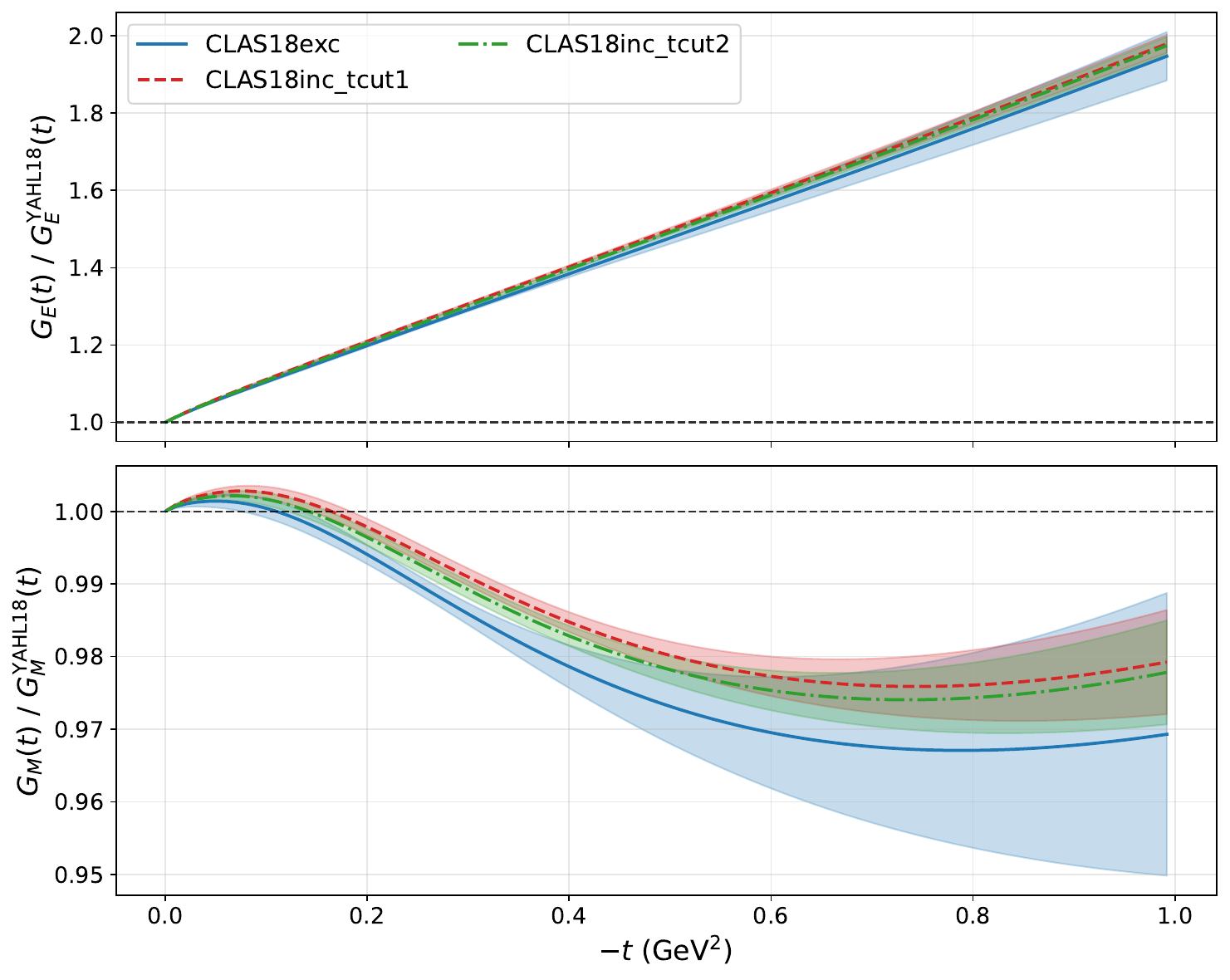}   
\caption{Comparison of the Sachs electric and magnetic form factors, $G_E(t)$ (upper panel) and $G_M(t)$ (lower panel), obtained from the CLAS18exc, CLAS18inc\_tcut1, and CLAS18inc\_tcut2 analyses in which the Pauli form factor $F_2(t)$ is fixed to the Kelly parametrization~\cite{Kelly:2004hm}, while the Dirac form factor $F_1(t)$ is described by the $P$-pole parametrization of Eq.~(\ref{Eq8}). The results are presented as ratios to the YAHL18 parametrization of Ref.~\cite{Ye:2017gyb}, extracted from a global analysis of elastic electron--proton scattering data.}
\label{fig:FixedF2}
\end{figure}

Figure~\ref{fig:FixedF2} compares the Sachs form factors obtained from the CLAS18exc, CLAS18inc\_tcut1 ($|t|_{\rm cut}=0.265~\mathrm{GeV}^2$), and CLAS18inc\_tcut2 ($|t|_{\rm cut}=0.343~\mathrm{GeV}^2$) analyses, in which $F_2(t)$ is fixed to the Kelly parametrization~\cite{Kelly:2004hm} and only $F_1(t)$ is described by the $P$-pole parametrization. As in the previous figures, the results are presented as ratios to the YAHL18 parametrization~\cite{Ye:2017gyb} to show the differences better. 
The electric form factor $G_E(t)$ shows essentially the same behavior as in the previous analyses so that all three fits are lying systematically above the YAHL18 parametrization and remain in excellent agreement with one another over the whole $|t|$ range. For the magnetic form factor $G_M(t)$, the differences between the three fits are reduced compared with the previous $P$-pole analyses. This is expected since $F_2(t)$ is fixed to the same Kelly parametrization in all three fits. The remaining differences in $G_M(t)$ originate from the fitted Dirac form factor $F_1(t)$, considering $ G_M(t)=F_1(t)+F_2(t) .$ 

As shown in Eq.~(\ref{Eq6}), the electromagnetic radii of the proton are determined from the slopes of the Sachs FFs $G_E(t)$ and $G_M(t)$ at $t=0$. 
Therefore, they are highly sensitive to the behavior of these FFs at small-$|t|$. 
Considering the differences observed between the various analyses performed in this section, we investigate in the next section how these differences affect the extracted charge and magnetic radii of the proton.

%

\section{The charge and magnetic radii}\label{sec:five} 

For years, the charge and magnetic radii of the nucleon which give us unique information about its spatial distribution of the electric charge and magnetization, have been interesting subjects
experimentally~\cite{Pohl:2010zza,Epstein:2014zua,Beyer:2017gug,Fleurbaey:2018fih,Mihovilovic:2019jiz,Bezginov:2019mdi,Karr:2020wgh,Heacock:2021btd,Gao:2021sml,Atac:2021wqj,Xiong:2023zih,Lin:2024bzo}, 
theoretically~\cite{Aliev:2008cs,Bauer:2012pv,Anikin:2013aka,Sharma:2013mfa,Flores-Mendieta:2015wir,He:2017viu,Sufian:2016hwn,QCDSF:2017ssq,Hasan:2017wwt,HillerBlin:2017syu,Alexandrou:2018sjm,Alarcon:2018zbz,Shintani:2018ozy,Filin:2019eoe,Mondal:2019jdg,Jang:2019jkn,Barabanov:2020jvn,Yang:2020rpi,Park:2021ypf,Contreras:2021epz,Mamo:2021jhj,Bar:2021crj,Xu:2021wwj,Brodsky:2022kef,Ding:2022ows,Djukanovic:2023beb,Chen:2023dxp,Alvarado:2023loi,Tsuji:2023llh,Kuzmin:2024ozz,Yao:2024uej}, 
and phenomenologically~\cite{Miller:2007uy,Hill:2010yb,Griffioen:2015hta,Lorenz:2014vha,Graczyk:2014lba,Higinbotham:2015rja,Lee:2015jqa,Horbatsch:2016ilr,Sick:2017aor,Sick:2018fzn,Horbatsch:2019wdn,Borah:2020gte,Alarcon:2020kcz,Atac:2020hdq,Gramolin:2021gln,Cui:2021vgm,Lin:2021xrc,Goharipour:2024mbk,Albloushi:2026omq,Qattan:2026piw,Krishna:2026yfh}, 
especially in the context of the proton radius puzzle~\cite{Barger:2010aj,Jentschura:2010ha,Carlson:2012pc,Lorenz:2012tm,Karr:2012mfa,Pohl:2013yb,Wang:2013fma,Onofrio:2013fea,Kraus:2014qua,Bernauer:2014cwa,Carlson:2015jba,Hill:2017wzi,Miller:2018ybm,Pacetti:2021fji,Peset:2021iul,Lin:2023fhr,Dahia:2023urs,Lumpay:2025btu}, which is attributed to the discrepancies observed between measurements obtained from electron scattering and laser spectroscopy of hydrogen atoms. For a recent review see Ref.~\cite{Goharipour:2025yxm} and references therein.

Having determined the proton electromagnetic FFs within the different fitting scenarios considered in this work, we now extract the corresponding charge and magnetic radii. As discussed in Sec.~\ref{sec:three}, the proton charge radius, $r_E$, and magnetic radius, $r_M$, can be calculated from the slopes of the Sachs FFs at $t=0$ according to Eq.~(\ref{Eq6}). Since the fitted parametrizations provide analytic expressions for $G_E(t)$ and $G_M(t)$, the radii can be evaluated directly from the corresponding derivatives at the origin. The determination of the proton radii provides an important consistency check of the extracted FFs and allows a direct comparison with previous determinations based on elastic electron--proton scattering, muonic hydrogen spectroscopy, recent global analyses, and theoretical calculations. More importantly, it enables us to assess how the extraction of electromagnetic FFs from BH-dominated EP measurements contributes to the ongoing discussion of the proton charge-radius puzzle.

\begin{figure}[!tb]
    \centering
\includegraphics[scale=1.0]{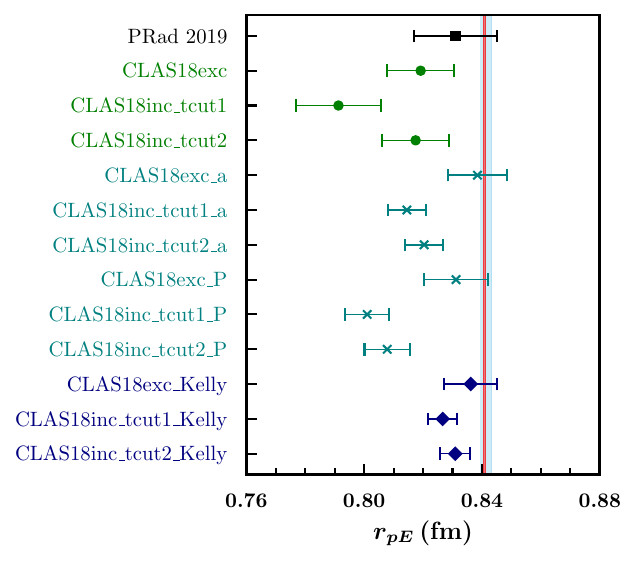}   
\caption{Comparison of the proton charge radius $r_E$ extracted from the various fitting scenarios considered in the present work. The results obtained using the dipole, $P$-pole, and fixed-$F_2$ parametrizations are represented by circles, crosses, and diamonds, respectively. The square symbol denotes the value measured by the PRad experiment~\cite{Xiong:2019umf}. The narrow and wide vertical bands indicate the PDG average~\cite{ParticleDataGroup:2024cfk} and the CODATA recommended value~\cite{Tiesinga:2021myr}, respectively.}
\label{fig:rEp}
\end{figure}

Figure~\ref{fig:rEp} compares the values of the proton charge radius, $r_E$, extracted from the electric form factor $G_E(t)$ for the various fitting scenarios considered in this work. For comparison, we also includes the PRad measurement~\cite{Xiong:2019umf}, shown by the square symbol, together with the PDG average~\cite{ParticleDataGroup:2024cfk} and the CODATA recommended value~\cite{Tiesinga:2021myr}, represented by the narrow and wide vertical bands, respectively. The results obtained using the dipole, $P$-pole, and fixed-$F_2$ parametrizations are denoted by circles, crosses, and diamonds, respectively.
As can be seen, all fitting scenarios lead to values of $r_E$ that are systematically smaller than the PDG and CODATA values. Furthermore, including the full CLAS 2018 dataset in the analysis leads to an even smaller charge radius, though it reduces the final uncertainties. 

Overall, the results obtained here indicate that the BH-dominated EP measurements favor a relatively small proton charge radius. The extracted values are generally smaller than most determinations based on elastic electron--proton scattering~\cite{A1:2010nsl,Zhan:2011ji,Mihovilovic:2019jiz}, while remaining compatible, within uncertainties, with the PRad measurement~\cite{Xiong:2019umf}, muonic hydrogen spectroscopy~\cite{Pohl:2010zza,Antognini:2013txn}, recent lattice QCD calculations~\cite{Jang:2019jkn,Alexandrou:2020aja,Tsuji:2023llh}, and holographic QCD predictions~\cite{Mamo:2021jhj,Ahmady:2021qed,Xu:2021wwj}. By contrast, most phenomenological analyses based on elastic scattering data tend to predict larger values of $r_E$. For reference, the average values obtained from lattice QCD, theoretical studies, and phenomenological analyses are $0.815 \pm 0.016$ fm, $0.827 \pm 0.013$ fm, and $0.857 \pm 0.002$ fm, respectively~\cite{Goharipour:2025yxm}.

\begin{figure}[!tb]
    \centering
\includegraphics[scale=1.0]{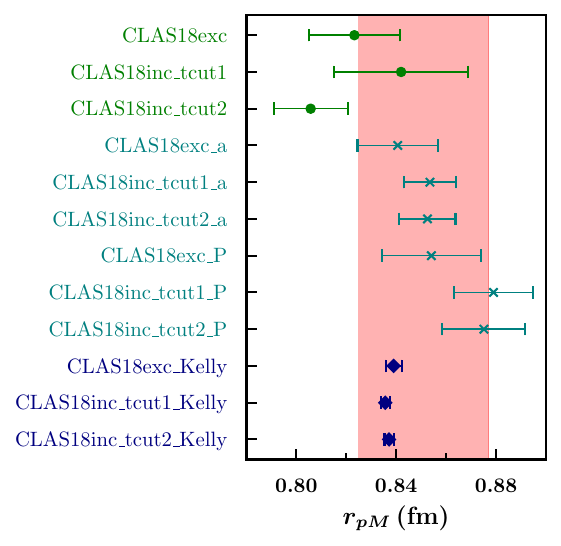}   
\caption{Comparison of the proton magnetic radius $r_M$ extracted from the various fitting scenarios considered in the present work. The results obtained using the dipole, $P$-pole, and fixed-$F_2$ parametrizations are represented by circles, crosses, and diamonds, respectively. The vertical band denotes the PDG average~\cite{ParticleDataGroup:2024cfk}.}
\label{fig:rMp}
\end{figure}

Figure~\ref{fig:rMp} compares the proton magnetic radius, $r_M$, extracted from the various fitting scenarios considered in this work with the corresponding value reported by PDG, represented by the vertical band. In contrast to the charge radius, all extracted values of $r_M$ are almost consistent within uncertainties and are also compatible with the PDG average, $r_M = 0.851 \pm 0.026$ fm~\cite{ParticleDataGroup:2024cfk}.
Unlike the case of the charge radius, no clear systematic effect is observed from the inclusion of the CLAS 2018 dataset. Depending on the fitting scenario, the inclusion of these data may either increase or decrease the central value of $r_M$. As expected, the fixed-$F_2$ analysis leads to the smallest uncertainties which reflects the fact that the magnetic form factor is much better constrained once the Pauli form factor is fixed to the Kelly parametrization.

Overall, our results demonstrate that, in kinematic regions where the BH process dominates the cross section, the EP measurements  provide a viable and complementary approach for studying the proton electromagnetic structure. Particularly, they have potential to show accurate constraints on the electromagnetic FFs, especially the Dirac form
factor. Such measurements, especially at smaller values
of $ t $, can also shed light on the long-standing charge-radius puzzle.
According to the results obtained in this section, one can conclude that he extracted electromagnetic FFs consistently favor a relatively small proton charge radius.

%

\section{Summary and conclusion}\label{sec:six} 

In this work, we extended our previous study on the proton electromagnetic FFs extraction from EP measurements in the kinematic region where the BH contribution is dominant in the cross section~\cite{Moradi:2025pkp}. Unlike our earlier study, which included just the CLAS 2015 data~\cite{CLAS:2015uuo}, the present analysis includes essentially all available high-precision EP cross section measurements from CLAS and Hall~A Collaborations. This enlarged dataset allows us to perform the first global analysis of BH-dominated EP measurements to extract the proton Sachs FFs and their corresponding charge and magnetic radii.

To this aim, we adopted the same phenomenological framework developed in Ref.~\cite{Moradi:2025pkp}, where the BH contribution provides direct sensitivity to the proton Dirac and Pauli FFs, $F_1(t)$ and $F_2(t)$. We examined both dipole and $P$-pole parametrizations of the FFs and considered several fitting strategies, including analyses in which the Pauli form factor was fixed to the Kelly parametrization~\cite{Kelly:2004hm} considering the fact that the EP measurements provide strong constraints on $F_1(t)$, whereas the sensitivity to $F_2(t)$ is limited. As a result, we found that the EP measurements lead to larger values of the electric form factor $G_E(t)$ compared with the YAHL18 parametrization~\cite{Ye:2017gyb}, which is based on a global analysis of elastic electron--proton scattering data. The differences become more and more as $|t|$ increases. In contrast, the magnetic form factor $G_M(t)$ is affected much less so that the deviations generally do not exceed about $7\%$.

A detailed comparison of the individual datasets reveals a significant tension between the CLAS 2018 measurements~\cite{CLAS:2018bgk} and the remaining EP data within the present phenomenological framework. When all datasets are included simultaneously, the fit yields a relatively large $\chi^{2}/\mathrm{d.o.f.}$, and the extracted FFs lead to unusual values of charge and large magnetic radii. By excluding the CLAS 2018 dataset or applying suitable low-$|t|$ cuts, we obtained stable fits with $\chi^{2}/\mathrm{d.o.f.}$ close to unity and FFs that exhibit good agreement among the different parametrizations. By further investigation, we showed that the improvement is not simply due to a reduction of the $\chi^2$ contribution from the CLAS 2018 data themselves; rather, the individual $\chi^2$ contributions of several of the other datasets also decrease as the low-$|t|$ CLAS 2018 points are excluded.

Using the extracted Sachs FFs, we determined the proton charge and magnetic radii for all fitting scenarios. For all analyses, the charge radius, $ r_E $, obtained is smaller than the values reported by PDG~\cite{ParticleDataGroup:2024cfk} and most determinations based on elastic electron--proton scattering. On the other hand, it is compatible, within uncertainties, with the PRad measurement~\cite{Xiong:2019umf}, muonic hydrogen spectroscopy, and several recent theoretical calculations, including lattice QCD. In contrast, we found that the extracted magnetic radius, $ r_M $, is consistent with the PDG average for all fitting scenarios considered in this work.

The present results demonstrate that BH-dominated EP measurements provide an independent and complementary probe of the proton electromagnetic structure. Although the current data cannot provide sufficient sensitivity to determine the Pauli form factor with high precision, they already offer meaningful constraints on the Dirac form factor. The main result of the present study is that the EP measurements favor a relatively small proton charge radius. This suggests that exclusive photon leptoproduction processes can provide valuable input to ongoing investigations of the proton charge-radius puzzle.
The approach developed in the present study will be strengthened by the
future high-precision EP measurements, particularly at lower values of $|t|$, where the proton radii are most sensitive to the form factor slopes. Moreover, the present work opens the possibility of performing combined analyses of EP and elastic electron--proton scattering data within a unified framework, which may provide the most reliable determination of the proton electromagnetic FFs and, consequently, of the charge and magnetic radii.

%

%

\bibliographystyle{apsrev4-1}
\bibliography{article}

@article{Diehl:2013xca,
    author = "Diehl, Markus and Kroll, Peter",
    title = "{Nucleon form factors, generalized parton distributions and quark angular momentum}",
    eprint = "1302.4604",
    archivePrefix = "arXiv",
    primaryClass = "hep-ph",
    reportNumber = "DESY-13-025, DESY 13-025",
    doi = "10.1140/epjc/s10052-013-2397-7",
    journal = "Eur. Phys. J. C",
    volume = "73",
    number = "4",
    pages = "2397",
    year = "2013"
}

@article{Hashamipour:2022noy,
    author = "Hashamipour, Hadi and Goharipour, Muhammad and Azizi, K. and Goloskokov, S. V.",
    title = "{Generalized parton distributions at zero skewness}",
    eprint = "2211.09522",
    archivePrefix = "arXiv",
    primaryClass = "hep-ph",
    doi = "10.1103/PhysRevD.107.096005",
    journal = "Phys. Rev. D",
    volume = "107",
    number = "9",
    pages = "096005",
    year = "2023"
}

@article{Ramalho:2023hqd,
    author = "Ramalho, G. and Pe{\~n}a, M. T.",
    title = "{Electromagnetic transition form factors of baryon resonances}",
    eprint = "2306.13900",
    archivePrefix = "arXiv",
    primaryClass = "hep-ph",
    doi = "10.1016/j.ppnp.2024.104097",
    journal = "Prog. Part. Nucl. Phys.",
    volume = "136",
    pages = "104097",
    year = "2024"
}

@article{Pate:2024acz,
    author = "Pate, S. F. and Papavassiliou, V. and Schaub, J. P. and Trujillo, D. P. and Ivanov, M. V. and Barbaro, M. B. and Giusti, C.",
    title = "{Global fit of electron and neutrino elastic scattering data to determine the strange quark contribution to the vector and axial form factors of the nucleon}",
    eprint = "2402.10854",
    archivePrefix = "arXiv",
    primaryClass = "hep-ph",
    reportNumber = "FERMILAB-PUB-24-0074-V",
    doi = "10.1103/PhysRevD.109.093001",
    journal = "Phys. Rev. D",
    volume = "109",
    number = "9",
    pages = "093001",
    year = "2024"
}

@article{Yao:2024uej,
    author = "Yao, Zhao-Qian and Binosi, Daniele and Cui, Zhu-Fang and Roberts, Craig D.",
    title = "{Nucleon charge and magnetisation distributions: Flavour separation and zeroes}",
    eprint = "2403.08088",
    archivePrefix = "arXiv",
    primaryClass = "hep-ph",
    reportNumber = "NJU-INP 085/24",
    doi = "10.1016/j.fmre.2024.11.005",
    journal = "Fund. Res.",
    volume = "6",
    pages = "1409--1416",
    year = "2026"
}

@article{Wang:2024abv,
    author = "Wang, Jiaqi and Fu, Dongyan and Dong, Yubing",
    title = "{A systematic study of nucleon form factors with the pion cloud effect}",
    eprint = "2410.14953",
    archivePrefix = "arXiv",
    primaryClass = "hep-ph",
    doi = "10.1140/epjc/s10052-025-14908-1",
    journal = "Eur. Phys. J. C",
    volume = "85",
    number = "11",
    pages = "1254",
    year = "2025"
}

@article{Hernandez-Pinto:2024kwg,
    author = "Hern{\'a}ndez-Pinto, R. J. and Guti{\'e}rrez-Guerrero, L. X. and Bedolla, M. A. and Bashir, A.",
    title = "{Electric, magnetic, and quadrupole form factors and charge radii of vector mesons: From light to heavy sectors in a contact interaction}",
    eprint = "2410.23813",
    archivePrefix = "arXiv",
    primaryClass = "hep-ph",
    doi = "10.1103/PhysRevD.110.114015",
    journal = "Phys. Rev. D",
    volume = "110",
    number = "11",
    pages = "114015",
    year = "2024"
}

@article{Cheng:2024cxk,
    author = "Cheng, Peng and Yao, Zhao-Qian and Binosi, Daniele and Roberts, Craig D.",
    title = "{Likelihood of a zero in the proton elastic electric form factor}",
    eprint = "2412.10598",
    archivePrefix = "arXiv",
    primaryClass = "hep-ph",
    reportNumber = "NJU-INP 095/24",
    doi = "10.1016/j.physletb.2025.139323",
    journal = "Phys. Lett. B",
    volume = "862",
    pages = "139323",
    year = "2025"
}

@inbook{Lin:2024rak,
    author = "Lin, Yong-Hui and Hammer, Hans-Werner and Mei{\ss}ner, Ulf-G.",
    editor = "Fischer, Christian S. and Guo, Feng-Kun and Pich, Antonio",
    title = "{Baryon Form Factors}",
    booktitle = "{Encyclopedia of Particle Physics: Volume 2}",
    eprint = "2412.12885",
    archivePrefix = "arXiv",
    primaryClass = "hep-ph",
    doi = "10.1016/B978-0-443-26598-3.00022-5",
    publisher = "Elsevier",
    volume = "2",
    pages = "148--166",
    month = "7",
    year = "2026"
}

@article{Kuzmin:2024ozz,
    author = "Kuzmin, K. S. and Levashko, N. M. and Krivoruchenko, M. I.",
    title = "{Electromagnetic nucleon form factors in the extended vector meson dominance model}",
    eprint = "2412.13150",
    archivePrefix = "arXiv",
    primaryClass = "hep-ph",
    doi = "10.1103/PhysRevD.111.013004",
    journal = "Phys. Rev. D",
    volume = "111",
    number = "1",
    pages = "013004",
    year = "2025"
}

@article{Cheng:2025yij,
    author = "Cheng, Peng and Yao, Zhao Qian and Binosi, Daniele and Lu, Ya and Roberts, Craig D.",
    title = "{Quark + diquark description of nucleon elastic electromagnetic form factors}",
    eprint = "2507.13484",
    archivePrefix = "arXiv",
    primaryClass = "hep-ph",
    reportNumber = "NJU-INP 104/25",
    doi = "10.1140/epja/s10050-025-01724-0",
    journal = "Eur. Phys. J. A",
    volume = "61",
    number = "11",
    pages = "255",
    year = "2025"
}

@article{Alexandrou:2025vto,
    author = "Alexandrou, Constantia and Bacchio, Simone and Koutsou, Giannis and Prasad, Bhavna and Spanoudes, Gregoris",
    title = "{Proton and neutron electromagnetic form factors in the continuum limit using lattice QCD ensembles with physical pion masses}",
    eprint = "2507.20910",
    archivePrefix = "arXiv",
    primaryClass = "hep-lat",
    doi = "10.1103/tt39-n1df",
    journal = "Phys. Rev. D",
    volume = "113",
    number = "11",
    pages = "114524",
    year = "2026"
}

@article{Yu:2025jcs,
    author = "Yu, Ji-Xin and Cheng, Shan and Han, Jia-Jie and Li, Hsiang-nan and Yu, Fu-Sheng",
    title = "{Tomography of high-twist proton structure through ep elastic scattering}",
    eprint = "2511.12589",
    archivePrefix = "arXiv",
    primaryClass = "hep-ph",
    doi = "10.1140/epjc/s10052-026-15554-x",
    journal = "Eur. Phys. J. C",
    volume = "86",
    number = "3",
    pages = "300",
    year = "2026"
}

@article{Williams:2025fiv,
    author = "Williams, Tyler and Rittenhouse West, Jennifer and Higinbotham, Douglas W. and Benmokhtar, Fatiha",
    title = "{Hofstadter-Herman Visualization as a Diagnostic Tool for Systematic Effects in Electromagnetic Form Factor Extractions}",
    eprint = "2511.12007",
    archivePrefix = "arXiv",
    primaryClass = "nucl-ex",
    month = "11",
    year = "2025"
}

@article{Arbabifar:2026tev,
    author = "Arbabifar, Fatemeh and Morshedian, Nader and Atashbar Tehrani, Shahin",
    title = "{New analysis for nucleon form factors from GPDs}",
    eprint = "2605.14083",
    archivePrefix = "arXiv",
    primaryClass = "hep-ph",
    doi = "10.1140/epjp/s13360-026-08025-5",
    journal = "Eur. Phys. J. Plus",
    volume = "141",
    number = "8",
    pages = "927",
    year = "2026"
}

@article{Lee:2026zgo,
    author = "Lee, Hui-Jae and Choi, Yongwoo and Kim, Hyun-Chul",
    title = "{Electromagnetic form factors of the nucleon from the instanton vacuum}",
    eprint = "2605.25900",
    archivePrefix = "arXiv",
    primaryClass = "hep-ph",
    reportNumber = "INHA-NTG-03/2026",
    doi = "10.1103/mwlt-rbqh",
    journal = "Phys. Rev. D",
    volume = "114",
    number = "5",
    pages = "054006",
    year = "2026"
}

@article{Pohl:2010zza,
    author = "Pohl, Randolf and others",
    title = "{The size of the proton}",
    doi = "10.1038/nature09250",
    journal = "Nature",
    volume = "466",
    pages = "213--216",
    year = "2010"
}

@article{Lorenz:2012tm,
    author = "Lorenz, I. T. and Hammer, H. -W. and Meissner, Ulf-G.",
    title = "{The size of the proton - closing in on the radius puzzle}",
    eprint = "1205.6628",
    archivePrefix = "arXiv",
    primaryClass = "hep-ph",
    doi = "10.1140/epja/i2012-12151-1",
    journal = "Eur. Phys. J. A",
    volume = "48",
    pages = "151",
    year = "2012"
}

@article{Pohl:2013yb,
    author = "Pohl, Randolf and Gilman, Ronald and Miller, Gerald A. and Pachucki, Krzysztof",
    title = "{Muonic hydrogen and the proton radius puzzle}",
    eprint = "1301.0905",
    archivePrefix = "arXiv",
    primaryClass = "physics.atom-ph",
    doi = "10.1146/annurev-nucl-102212-170627",
    journal = "Ann. Rev. Nucl. Part. Sci.",
    volume = "63",
    pages = "175--204",
    year = "2013"
}

@article{Carlson:2015jba,
    author = "Carlson, Carl E.",
    title = "{The Proton Radius Puzzle}",
    eprint = "1502.05314",
    archivePrefix = "arXiv",
    primaryClass = "hep-ph",
    doi = "10.1016/j.ppnp.2015.01.002",
    journal = "Prog. Part. Nucl. Phys.",
    volume = "82",
    pages = "59--77",
    year = "2015"
}

@article{Karr:2020wgh,
    author = "Karr, Jean-Philippe and Marchand, Dominique and Voutier, Eric",
    title = "{The proton size}",
    doi = "10.1038/s42254-020-0229-x",
    journal = "Nature Rev. Phys.",
    volume = "2",
    number = "11",
    pages = "601--614",
    year = "2020"
}

@article{Goharipour:2025yxm,
    author = "Goharipour, Muhammad and Irani, Fatemeh and Amiri, M. H. and Fatehi, H. and Falahi, Behnam and Moradi, A. and Azizi, K.",
    collaboration = "MMGPDs",
    title = "{Can we determine the exact size of the nucleon?: A comprehensive study of different radii}",
    eprint = "2503.08847",
    archivePrefix = "arXiv",
    primaryClass = "hep-ph",
    doi = "10.1016/j.nuclphysb.2025.116962",
    journal = "Nucl. Phys. B",
    volume = "1017",
    pages = "116962",
    year = "2025"
}

@article{Kelly:2004hm,
    author = "Kelly, J. J.",
    title = "{Simple parametrization of nucleon form factors}",
    doi = "10.1103/PhysRevC.70.068202",
    journal = "Phys. Rev. C",
    volume = "70",
    pages = "068202",
    year = "2004"
}

@article{Qattan:2004ht,
    author = "Qattan, I. A. and others",
    title = "{Precision Rosenbluth measurement of the proton elastic form-factors}",
    eprint = "nucl-ex/0410010",
    archivePrefix = "arXiv",
    reportNumber = "JLAB-PHY-04-33",
    doi = "10.1103/PhysRevLett.94.142301",
    journal = "Phys. Rev. Lett.",
    volume = "94",
    pages = "142301",
    year = "2005"
}

@article{Perdrisat:2006hj,
    author = "Perdrisat, C. F. and Punjabi, V. and Vanderhaeghen, M.",
    title = "{Nucleon Electromagnetic Form Factors}",
    eprint = "hep-ph/0612014",
    archivePrefix = "arXiv",
    reportNumber = "WM-06-115, JLAB-THY-06-595",
    doi = "10.1016/j.ppnp.2007.05.001",
    journal = "Prog. Part. Nucl. Phys.",
    volume = "59",
    pages = "694--764",
    year = "2007"
}

@article{Crawford:2006rz,
    author = "Crawford, Christopher B. and others",
    title = "{Measurement of the proton electric to magnetic form factor ratio from vector H-1(vector e, e' p)}",
    eprint = "nucl-ex/0609007",
    archivePrefix = "arXiv",
    doi = "10.1103/PhysRevLett.98.052301",
    journal = "Phys. Rev. Lett.",
    volume = "98",
    pages = "052301",
    year = "2007"
}

@article{Arrington:2007ux,
    author = "Arrington, J. and Melnitchouk, W. and Tjon, J. A.",
    title = "{Global analysis of proton elastic form factor data with two-photon exchange corrections}",
    eprint = "0707.1861",
    archivePrefix = "arXiv",
    primaryClass = "nucl-ex",
    reportNumber = "JLAB-THY-07-678",
    doi = "10.1103/PhysRevC.76.035205",
    journal = "Phys. Rev. C",
    volume = "76",
    pages = "035205",
    year = "2007"
}

@article{A1:2010nsl,
    author = "Bernauer, J. C. and others",
    collaboration = "A1",
    title = "{High-precision determination of the electric and magnetic form factors of the proton}",
    eprint = "1007.5076",
    archivePrefix = "arXiv",
    primaryClass = "nucl-ex",
    doi = "10.1103/PhysRevLett.105.242001",
    journal = "Phys. Rev. Lett.",
    volume = "105",
    pages = "242001",
    year = "2010"
}

@article{A1:2013fsc,
    author = "Bernauer, J. C. and others",
    collaboration = "A1",
    title = "{Electric and magnetic form factors of the proton}",
    eprint = "1307.6227",
    archivePrefix = "arXiv",
    primaryClass = "nucl-ex",
    doi = "10.1103/PhysRevC.90.015206",
    journal = "Phys. Rev. C",
    volume = "90",
    number = "1",
    pages = "015206",
    year = "2014"
}

@article{Zhan:2011ji,
    author = "Zhan, X. and others",
    title = "{High-Precision Measurement of the Proton Elastic Form Factor Ratio $\mu_pG_E/G_M$ at low $Q^2$}",
    eprint = "1102.0318",
    archivePrefix = "arXiv",
    primaryClass = "nucl-ex",
    reportNumber = "JLAB-PHY-11-1311",
    doi = "10.1016/j.physletb.2011.10.002",
    journal = "Phys. Lett. B",
    volume = "705",
    pages = "59--64",
    year = "2011"
}

@article{Mihovilovic:2016rkr,
    author = "Mihovilovi{\v{c}}, M. and others",
    title = "{First measurement of proton's charge form factor at very low $Q^2$ with initial state radiation}",
    eprint = "1612.06707",
    archivePrefix = "arXiv",
    primaryClass = "nucl-ex",
    doi = "10.1016/j.physletb.2017.05.031",
    journal = "Phys. Lett. B",
    volume = "771",
    pages = "194--198",
    year = "2017"
}

@article{Ye:2017gyb,
    author = "Ye, Zhihong and Arrington, John and Hill, Richard J. and Lee, Gabriel",
    title = "{Proton and Neutron Electromagnetic Form Factors and Uncertainties}",
    eprint = "1707.09063",
    archivePrefix = "arXiv",
    primaryClass = "nucl-ex",
    reportNumber = "FERMILAB-PUB-17-281-T",
    doi = "10.1016/j.physletb.2017.11.023",
    journal = "Phys. Lett. B",
    volume = "777",
    pages = "8--15",
    year = "2018"
}

@article{Xiong:2019umf,
    author = "Xiong, W. and others",
    title = "{A small proton charge radius from an electron{\textendash}proton scattering experiment}",
    doi = "10.1038/s41586-019-1721-2",
    journal = "Nature",
    volume = "575",
    number = "7781",
    pages = "147--150",
    year = "2019"
}

@article{Lin:2021umz,
    author = "Lin, Yong-Hui and Hammer, Hans-Werner and Mei{\ss}ner, Ulf-G.",
    title = "{Dispersion-theoretical analysis of the electromagnetic form factors of the nucleon: Past, present and future}",
    eprint = "2106.06357",
    archivePrefix = "arXiv",
    primaryClass = "hep-ph",
    doi = "10.1140/epja/s10050-021-00562-0",
    journal = "Eur. Phys. J. A",
    volume = "57",
    number = "8",
    pages = "255",
    year = "2021"
}

@article{Christy:2021snt,
    author = "Christy, M. E. and others",
    title = "{Form Factors and Two-Photon Exchange in High-Energy Elastic Electron-Proton Scattering}",
    eprint = "2103.01842",
    archivePrefix = "arXiv",
    primaryClass = "nucl-ex",
    reportNumber = "JLAB-PHY-21-3336",
    doi = "10.1103/PhysRevLett.128.102002",
    journal = "Phys. Rev. Lett.",
    volume = "128",
    number = "10",
    pages = "102002",
    year = "2022"
}

@article{Qattan:2024pco,
    author = "Qattan, I. A. and others",
    title = "{High precision measurements of the proton elastic electromagnetic form factors and their ratio at Q2=0.50,2.64,3.20, and 4.10GeV2}",
    eprint = "2411.05201",
    archivePrefix = "arXiv",
    primaryClass = "nucl-ex",
    doi = "10.1103/4vmq-s4c7",
    journal = "Phys. Rev. C",
    volume = "112",
    number = "3",
    pages = "035205",
    year = "2025"
}

@article{Xu:2021wwj,
    author = "Xu, Siqi and Mondal, Chandan and Lan, Jiangshan and Zhao, Xingbo and Li, Yang and Vary, James P.",
    collaboration = "BLFQ",
    title = "{Nucleon structure from basis light-front quantization}",
    eprint = "2108.03909",
    archivePrefix = "arXiv",
    primaryClass = "hep-ph",
    doi = "10.1103/PhysRevD.104.094036",
    journal = "Phys. Rev. D",
    volume = "104",
    number = "9",
    pages = "094036",
    year = "2021"
}

@article{Park:2021ypf,
    author = "Park, Sungwoo and Gupta, Rajan and Yoon, Boram and Mondal, Santanu and Bhattacharya, Tanmoy and Jang, Yong-Chull and Jo{\'o}, B{\'a}lint and Winter, Frank",
    collaboration = "Nucleon Matrix Elements (NME)",
    title = "{Precision nucleon charges and form factors using (2+1)-flavor lattice QCD}",
    eprint = "2103.05599",
    archivePrefix = "arXiv",
    primaryClass = "hep-lat",
    reportNumber = "LA-UR-21-20526, JLAB-THY-22-3583",
    doi = "10.1103/PhysRevD.105.054505",
    journal = "Phys. Rev. D",
    volume = "105",
    number = "5",
    pages = "054505",
    year = "2022"
}

@article{Goharipour:2024mbk,
    author = "Goharipour, Muhammad and Irani, Fatemeh and Hashamipour, Hadi and Azizi, K.",
    collaboration = "MMGPDs",
    title = "{The charge and magnetic radii of the nucleons from the generalized parton distributions}",
    eprint = "2408.01783",
    archivePrefix = "arXiv",
    primaryClass = "hep-ph",
    doi = "10.1016/j.physletb.2025.139423",
    journal = "Phys. Lett. B",
    volume = "864",
    pages = "139423",
    year = "2025"
}

@article{Ji:1996nm,
    author = "Ji, Xiang-Dong",
    title = "{Deeply virtual Compton scattering}",
    eprint = "hep-ph/9609381",
    archivePrefix = "arXiv",
    reportNumber = "UMD-PP-97-26, MIT-CTP-2568",
    doi = "10.1103/PhysRevD.55.7114",
    journal = "Phys. Rev. D",
    volume = "55",
    pages = "7114--7125",
    year = "1997"
}

@article{Collins:1998be,
    author = "Collins, John C. and Freund, Andreas",
    title = "{Proof of factorization for deeply virtual Compton scattering in QCD}",
    eprint = "hep-ph/9801262",
    archivePrefix = "arXiv",
    reportNumber = "PSU-TH-192",
    doi = "10.1103/PhysRevD.59.074009",
    journal = "Phys. Rev. D",
    volume = "59",
    pages = "074009",
    year = "1999"
}

@article{Goeke:2001tz,
    author = "Goeke, K. and Polyakov, Maxim V. and Vanderhaeghen, M.",
    title = "{Hard exclusive reactions and the structure of hadrons}",
    eprint = "hep-ph/0106012",
    archivePrefix = "arXiv",
    doi = "10.1016/S0146-6410(01)00158-2",
    journal = "Prog. Part. Nucl. Phys.",
    volume = "47",
    pages = "401--515",
    year = "2001"
}

@article{Belitsky:2001ns,
    author = "Belitsky, Andrei V. and Mueller, Dieter and Kirchner, A.",
    title = "{Theory of deeply virtual Compton scattering on the nucleon}",
    eprint = "hep-ph/0112108",
    archivePrefix = "arXiv",
    reportNumber = "DOE-ER-40762-009, UMD-PP-02-011, YITP-SB-01-51",
    doi = "10.1016/S0550-3213(02)00144-X",
    journal = "Nucl. Phys. B",
    volume = "629",
    pages = "323--392",
    year = "2002"
}

@article{Belitsky:2010jw,
    author = "Belitsky, A. V. and Mueller, Dieter",
    title = "{Exclusive electroproduction revisited: treating kinematical effects}",
    eprint = "1005.5209",
    archivePrefix = "arXiv",
    primaryClass = "hep-ph",
    doi = "10.1103/PhysRevD.82.074010",
    journal = "Phys. Rev. D",
    volume = "82",
    pages = "074010",
    year = "2010"
}

@article{Kriesten:2020wcx,
    author = "Kriesten, Brandon and Liuti, Simonetta",
    title = "{Theory of deeply virtual Compton scattering off the unpolarized proton}",
    eprint = "2004.08890",
    archivePrefix = "arXiv",
    primaryClass = "hep-ph",
    doi = "10.1103/PhysRevD.105.016015",
    journal = "Phys. Rev. D",
    volume = "105",
    number = "1",
    pages = "016015",
    year = "2022"
}

@article{Radyushkin:1997ki,
    author = "Radyushkin, A. V.",
    title = "{Nonforward parton distributions}",
    eprint = "hep-ph/9704207",
    archivePrefix = "arXiv",
    reportNumber = "JLAB-THY-97-10",
    doi = "10.1103/PhysRevD.56.5524",
    journal = "Phys. Rev. D",
    volume = "56",
    pages = "5524--5557",
    year = "1997"
}

@article{Diehl:2003ny,
    author = "Diehl, M.",
    title = "{Generalized parton distributions}",
    eprint = "hep-ph/0307382",
    archivePrefix = "arXiv",
    reportNumber = "DESY-THESIS-2003-018",
    doi = "10.1016/j.physrep.2003.08.002",
    journal = "Phys. Rept.",
    volume = "388",
    pages = "41--277",
    year = "2003"
}

@article{Ji:2004gf,
    author = "Ji, X.",
    title = "{Generalized parton distributions}",
    doi = "10.1146/annurev.nucl.54.070103.181302",
    journal = "Ann. Rev. Nucl. Part. Sci.",
    volume = "54",
    pages = "413--450",
    year = "2004"
}

@article{Belitsky:2005qn,
    author = "Belitsky, A. V. and Radyushkin, A. V.",
    title = "{Unraveling hadron structure with generalized parton distributions}",
    eprint = "hep-ph/0504030",
    archivePrefix = "arXiv",
    reportNumber = "JLAB-THY-04-34",
    doi = "10.1016/j.physrep.2005.06.002",
    journal = "Phys. Rept.",
    volume = "418",
    pages = "1--387",
    year = "2005"
}

@article{Boffi:2007yc,
    author = "Boffi, Sigfrido and Pasquini, Barbara",
    title = "{Generalized parton distributions and the structure of the nucleon}",
    eprint = "0711.2625",
    archivePrefix = "arXiv",
    primaryClass = "hep-ph",
    doi = "10.1393/ncr/i2007-10025-7",
    journal = "Riv. Nuovo Cim.",
    volume = "30",
    number = "9",
    pages = "387--448",
    year = "2007"
}

@article{Guidal:2013rya,
    author = "Guidal, Michel and Moutarde, Herv{\'e} and Vanderhaeghen, Marc",
    title = "{Generalized Parton Distributions in the valence region from Deeply Virtual Compton Scattering}",
    eprint = "1303.6600",
    archivePrefix = "arXiv",
    primaryClass = "hep-ph",
    doi = "10.1088/0034-4885/76/6/066202",
    journal = "Rept. Prog. Phys.",
    volume = "76",
    pages = "066202",
    year = "2013"
}

@article{Diehl:2015uka,
    author = "Diehl, Markus",
    title = "{Introduction to GPDs and TMDs}",
    eprint = "1512.01328",
    archivePrefix = "arXiv",
    primaryClass = "hep-ph",
    reportNumber = "DESY-15-234",
    doi = "10.1140/epja/i2016-16149-3",
    journal = "Eur. Phys. J. A",
    volume = "52",
    number = "6",
    pages = "149",
    year = "2016"
}

@article{Kumericki:2016ehc,
    author = "Kumericki, Kresimir and Liuti, Simonetta and Moutarde, Herve",
    title = "{GPD phenomenology and DVCS fitting}: {Entering the high-precision era}",
    eprint = "1602.02763",
    archivePrefix = "arXiv",
    primaryClass = "hep-ph",
    doi = "10.1140/epja/i2016-16157-3",
    journal = "Eur. Phys. J. A",
    volume = "52",
    number = "6",
    pages = "157",
    year = "2016"
}

@article{Mezrag:2022pqk,
    author = "Mezrag, C{\'e}dric",
    title = "{An Introductory Lecture on Generalised Parton Distributions}",
    eprint = "2207.13584",
    archivePrefix = "arXiv",
    primaryClass = "hep-ph",
    doi = "10.1007/s00601-022-01765-x",
    journal = "Few Body Syst.",
    volume = "63",
    number = "3",
    pages = "62",
    year = "2022"
}

@article{Xie:2023xkz,
    author = "Xie, Gang and Kou, Wei and Fu, Qiang and Ye, Zhenyu and Chen, Xurong",
    title = "{Deeply virtual compton scattering at future electron-ion colliders}",
    eprint = "2306.02357",
    archivePrefix = "arXiv",
    primaryClass = "hep-ph",
    doi = "10.1140/epjc/s10052-023-12065-x",
    journal = "Eur. Phys. J. C",
    volume = "83",
    number = "10",
    pages = "900",
    year = "2023"
}

@article{Guo:2023ahv,
    author = "Guo, Yuxun and Ji, Xiangdong and Santiago, M. Gabriel and Shiells, Kyle and Yang, Jinghong",
    title = "{Generalized parton distributions through universal moment parameterization: non-zero skewness case}",
    eprint = "2302.07279",
    archivePrefix = "arXiv",
    primaryClass = "hep-ph",
    doi = "10.1007/JHEP05(2023)150",
    journal = "JHEP",
    volume = "05",
    pages = "150",
    year = "2023"
}

@inbook{Lorce:2025aqp,
    author = "Lorc{\'e}, C{\'e}dric and Metz, Andreas and Pasquini, Barbara and Schweitzer, Peter",
    editor = "Fischer, Christian S. and Guo, Feng-Kun and Pich, Antonio",
    title = "{Parton Distribution Functions and their Generalizations}",
    booktitle = "{Encyclopedia of Particle Physics: Volume 2}",
    eprint = "2507.12664",
    archivePrefix = "arXiv",
    primaryClass = "hep-ph",
    doi = "10.1016/B978-0-443-26598-3.00088-2",
    publisher = "Elsevier",
    volume = "2",
    pages = "167--218",
    month = "7",
    year = "2026"
}

@article{Boer:2025ixc,
    author = {Bo{\"e}r, M. and others},
    title = "{Three-dimensional imaging of hadrons with hard exclusive reactions: advances in experiment, theory, phenomenology, and lattice QCD}",
    eprint = "2512.15064",
    archivePrefix = "arXiv",
    primaryClass = "hep-ph",
    month = "12",
    year = "2025"
}

@article{Burkardt:2002hr,
    author = "Burkardt, Matthias",
    title = "{Impact parameter space interpretation for generalized parton distributions}",
    eprint = "hep-ph/0207047",
    archivePrefix = "arXiv",
    doi = "10.1142/S0217751X03012370",
    journal = "Int. J. Mod. Phys. A",
    volume = "18",
    pages = "173--208",
    year = "2003"
}

@article{Kaur:2023lun,
    author = "Kaur, Satvir and Xu, Siqi and Mondal, Chandan and Zhao, Xingbo and Vary, James P.",
    collaboration = "BLFQ",
    title = "{Spatial imaging of proton via leading-twist nonskewed GPDs with basis light-front quantization}",
    eprint = "2307.09869",
    archivePrefix = "arXiv",
    primaryClass = "hep-ph",
    doi = "10.1103/PhysRevD.109.014015",
    journal = "Phys. Rev. D",
    volume = "109",
    number = "1",
    pages = "014015",
    year = "2024"
}

@article{Cichy:2024afd,
    author = "Cichy, Krzysztof and Constantinou, Martha and Sznajder, Pawe{\l} and Wagner, Jakub",
    title = "{Nucleon tomography and total angular momentum of valence quarks from synergy between lattice QCD and elastic scattering data}",
    eprint = "2409.17955",
    archivePrefix = "arXiv",
    primaryClass = "hep-ph",
    doi = "10.1103/PhysRevD.110.114025",
    journal = "Phys. Rev. D",
    volume = "110",
    number = "11",
    pages = "114025",
    year = "2024"
}

@article{Polyakov:2018zvc,
    author = "Polyakov, Maxim V. and Schweitzer, Peter",
    title = "{Forces inside hadrons: pressure, surface tension, mechanical radius, and all that}",
    eprint = "1805.06596",
    archivePrefix = "arXiv",
    primaryClass = "hep-ph",
    doi = "10.1142/S0217751X18300259",
    journal = "Int. J. Mod. Phys. A",
    volume = "33",
    number = "26",
    pages = "1830025",
    year = "2018"
}

@article{Burkert:2018bqq,
    author = "Burkert, V. D. and Elouadrhiri, L. and Girod, F. X.",
    title = "{The pressure distribution inside the proton}",
    doi = "10.1038/s41586-018-0060-z",
    journal = "Nature",
    volume = "557",
    number = "7705",
    pages = "396--399",
    year = "2018"
}

@article{Goharipour:2025lep,
    author = "Goharipour, Muhammad and Hashamipour, Hadi and Fatehi, H. and Irani, Fatemeh and Azizi, K. and Goloskokov, S. V.",
    collaboration = "MMGPDs",
    title = "{Mechanical properties of the nucleon from the generalized parton distributions}",
    eprint = "2501.16257",
    archivePrefix = "arXiv",
    primaryClass = "hep-ph",
    doi = "10.1103/jkhv-6949",
    journal = "Phys. Rev. D",
    volume = "112",
    number = "1",
    pages = "014016",
    year = "2025"
}

@article{Martinez-Fernandez:2025jvk,
    author = "Mart{\'\i}nez-Fern{\'a}ndez, V{\'\i}ctor and Binosi, Daniele and Mezrag, C{\'e}dric and Yao, Zhao-Qian",
    title = "{Constraining the energy-momentum tensor through the DVCS dispersion relation beyond leading power}",
    eprint = "2509.06669",
    archivePrefix = "arXiv",
    primaryClass = "hep-ph",
    doi = "10.1103/rqnd-gvlq",
    journal = "Phys. Rev. D",
    volume = "113",
    number = "9",
    pages = "094003",
    year = "2026"
}

@article{Moradi:2025pkp,
    author = "Moradi, Anoushiravan and Goharipour, Muhammad and Fatehi, H. and Azizi, K.",
    collaboration = "MMGPDs",
    title = "{Determination of proton electromagnetic form factors from DVCS measurements}",
    eprint = "2512.06554",
    archivePrefix = "arXiv",
    primaryClass = "hep-ph",
    doi = "10.1103/c8q9-vxtn",
    journal = "Phys. Rev. D",
    volume = "113",
    number = "5",
    pages = "054034",
    year = "2026"
}

@article{Adams:2024pxw,
    author = "Adams, Douglas Q. and others",
    title = "{Likelihood and Correlation Analysis of Compton Form Factors for Deeply Virtual Exclusive Scattering on the Nucleon}",
    eprint = "2410.23469",
    archivePrefix = "arXiv",
    primaryClass = "hep-ph",
    month = "10",
    year = "2024"
}

@article{Kriesten:2019jep,
    author = "Kriesten, Brandon and Liuti, Simonetta and Calero-Diaz, Liliet and Keller, Dustin and Meyer, Andrew and Goldstein, Gary R. and Osvaldo Gonzalez-Hernandez, J.",
    title = "{Extraction of generalized parton distribution observables from deeply virtual electron proton scattering experiments}",
    eprint = "1903.05742",
    archivePrefix = "arXiv",
    primaryClass = "hep-ph",
    doi = "10.1103/PhysRevD.101.054021",
    journal = "Phys. Rev. D",
    volume = "101",
    number = "5",
    pages = "054021",
    year = "2020"
}

@article{CLAS:2015uuo,
    author = "Jo, H. S. and others",
    collaboration = "CLAS",
    title = "{Cross sections for the exclusive photon electroproduction on the proton and Generalized Parton Distributions}",
    eprint = "1504.02009",
    archivePrefix = "arXiv",
    primaryClass = "hep-ex",
    reportNumber = "JLAB-PHY-15-2037",
    doi = "10.1103/PhysRevLett.115.212003",
    journal = "Phys. Rev. Lett.",
    volume = "115",
    number = "21",
    pages = "212003",
    year = "2015"
}

@article{Burkert:2025gzu,
    author = "Burkert, V. D. and Camsonne, A. and Chatagnon, P. and Cichy, K. and Constantinou, M. and Dutrieux, H. and Higuera-Angulo, I. M. and Mezrag, C. and Richards, D. and Sznajder, P.",
    title = "{Open database for GPD analyses}",
    eprint = "2503.18152",
    archivePrefix = "arXiv",
    primaryClass = "hep-ph",
    reportNumber = "JLAB-PHY-25-4269",
    doi = "10.1140/epjc/s10052-025-14512-3",
    journal = "Eur. Phys. J. C",
    volume = "85",
    number = "8",
    pages = "838",
    year = "2025"
}

@article{Kumericki:2006xx,
    author = "Kumericki, K. and Mueller, Dieter and Passek-Kumericki, K. and Schafer, A.",
    title = "{Deeply virtual Compton scattering beyond next-to-leading order: the flavor singlet case}",
    eprint = "hep-ph/0605237",
    archivePrefix = "arXiv",
    doi = "10.1016/j.physletb.2007.02.071",
    journal = "Phys. Lett. B",
    volume = "648",
    pages = "186--194",
    year = "2007"
}

@article{Kumericki:2007sa,
    author = "Kumericki, K. and Mueller, Dieter and Passek-Kumericki, K.",
    title = "{Towards a fitting procedure for deeply virtual Compton scattering at next-to-leading order and beyond}",
    eprint = "hep-ph/0703179",
    archivePrefix = "arXiv",
    doi = "10.1016/j.nuclphysb.2007.10.029",
    journal = "Nucl. Phys. B",
    volume = "794",
    pages = "244--323",
    year = "2008"
}

@article{Kumericki:2009uq,
    author = "Kumeri{\v{c}}ki, Kresimir and Mueller, Dieter",
    title = "{Deeply virtual Compton scattering at small $x_B$ and the access to the GPD H}",
    eprint = "0904.0458",
    archivePrefix = "arXiv",
    primaryClass = "hep-ph",
    doi = "10.1016/j.nuclphysb.2010.07.015",
    journal = "Nucl. Phys. B",
    volume = "841",
    pages = "1--58",
    year = "2010"
}

@article{Cuic:2023mki,
    author = "{\v{C}}ui{\'c}, Marija and Duplan{\v{c}}i{\'c}, Goran and Kumeri{\v{c}}ki, Kre{\v{s}}imir and Passek-K., Kornelija",
    title = "{NLO corrections to the deeply virtual meson production revisited: impact on the extraction of generalized parton distributions}",
    eprint = "2310.13837",
    archivePrefix = "arXiv",
    primaryClass = "hep-ph",
    doi = "10.1007/JHEP12(2023)192",
    journal = "JHEP",
    volume = "12",
    pages = "192",
    year = "2023",
    note = "[Erratum: JHEP 02, 225 (2024)]"
}

@article{CLAS:2018bgk,
    author = "Hirlinger Saylor, N. and others",
    collaboration = "CLAS",
    title = "{Measurement of Unpolarized and Polarized Cross Sections for Deeply Virtual Compton Scattering on the Proton at Jefferson Laboratory with CLAS}",
    eprint = "1810.02110",
    archivePrefix = "arXiv",
    primaryClass = "hep-ex",
    reportNumber = "JLAB-PHY-18-2853",
    doi = "10.1103/PhysRevC.98.045203",
    journal = "Phys. Rev. C",
    volume = "98",
    number = "4",
    pages = "045203",
    year = "2018"
}

@article{JeffersonLabHallA:2015dwe,
    author = "Defurne, M. and others",
    collaboration = "Jefferson Lab Hall A",
    title = "{E00-110 experiment at Jefferson Lab Hall A: Deeply virtual Compton scattering off the proton at 6 GeV}",
    eprint = "1504.05453",
    archivePrefix = "arXiv",
    primaryClass = "nucl-ex",
    reportNumber = "IRFU-15-12, JLAB-PHY-15-2038",
    doi = "10.1103/PhysRevC.92.055202",
    journal = "Phys. Rev. C",
    volume = "92",
    number = "5",
    pages = "055202",
    year = "2015"
}

@article{Defurne:2017paw,
    author = "Defurne, M. and others",
    title = "{A glimpse of gluons through deeply virtual compton scattering on the proton}",
    eprint = "1703.09442",
    archivePrefix = "arXiv",
    primaryClass = "hep-ex",
    reportNumber = "JLAB-PHY-17-2492",
    doi = "10.1038/s41467-017-01819-3",
    journal = "Nature Commun.",
    volume = "8",
    number = "1",
    pages = "1408",
    year = "2017"
}

@article{Benali:2020vma,
    author = "Benali, M. and others",
    title = "{Deeply virtual Compton scattering off the neutron}",
    eprint = "2109.02076",
    archivePrefix = "arXiv",
    primaryClass = "hep-ph",
    doi = "10.1038/s41567-019-0774-3",
    journal = "Nature Phys.",
    volume = "16",
    number = "2",
    pages = "191--198",
    year = "2020"
}

@article{James:1975dr,
    author = "James, F. and Roos, M.",
    title = "{Minuit: A System for Function Minimization and Analysis of the Parameter Errors and Correlations}",
    reportNumber = "CERN-DD-75-20",
    doi = "10.1016/0010-4655(75)90039-9",
    journal = "Comput. Phys. Commun.",
    volume = "10",
    pages = "343--367",
    year = "1975"
}

@article{iminuit,
  author={Hans Dembinski and Piti Ongmongkolkul et al.},
  title={scikit-hep/iminuit},
  DOI={10.5281/zenodo.3949207},
  publisher={Zenodo},
  year={2020},
  month={Dec},
  url={https://doi.org/10.5281/zenodo.3949207}
}

@article{ParticleDataGroup:2024cfk,
    author = "Navas, S. and others",
    collaboration = "Particle Data Group",
    title = "{Review of particle physics}",
    doi = "10.1103/PhysRevD.110.030001",
    journal = "Phys. Rev. D",
    volume = "110",
    number = "3",
    pages = "030001",
    year = "2024"
}

@article{Epstein:2014zua,
    author = "Epstein, Zachary and Paz, Gil and Roy, Joydeep",
    title = "{Model independent extraction of the proton magnetic radius from electron scattering}",
    eprint = "1407.5683",
    archivePrefix = "arXiv",
    primaryClass = "hep-ph",
    reportNumber = "WSU-HEP-1401",
    doi = "10.1103/PhysRevD.90.074027",
    journal = "Phys. Rev. D",
    volume = "90",
    number = "7",
    pages = "074027",
    year = "2014"
}

@article{Beyer:2017gug,
    author = "Beyer, Axel and others",
    title = "{The Rydberg constant and proton size from atomic hydrogen}",
    doi = "10.1126/science.aah6677",
    journal = "Science",
    volume = "358",
    number = "6359",
    pages = "79--85",
    year = "2017"
}

@article{Fleurbaey:2018fih,
    author = "Fleurbaey, H{\'e}l{\`e}ne and Galtier, Sandrine and Thomas, Simon and Bonnaud, Marie and Julien, Lucile and Biraben, Fran{\c{c}}ois and Nez, Fran{\c{c}}ois and Abgrall, Michel and Gu{\'e}na, Jocelyne",
    title = "{New Measurement of the $1S-3S$ Transition Frequency of Hydrogen: Contribution to the Proton Charge Radius Puzzle}",
    eprint = "1801.08816",
    archivePrefix = "arXiv",
    primaryClass = "physics.atom-ph",
    doi = "10.1103/PhysRevLett.120.183001",
    journal = "Phys. Rev. Lett.",
    volume = "120",
    number = "18",
    pages = "183001",
    year = "2018"
}

@article{Mihovilovic:2019jiz,
    author = "Mihovilovi{\v{c}}, M. and others",
    title = "{The proton charge radius extracted from the initial-state radiation experiment at MAMI}",
    eprint = "1905.11182",
    archivePrefix = "arXiv",
    primaryClass = "nucl-ex",
    doi = "10.1140/epja/s10050-021-00414-x",
    journal = "Eur. Phys. J. A",
    volume = "57",
    number = "3",
    pages = "107",
    year = "2021"
}

@article{Bezginov:2019mdi,
    author = "Bezginov, N. and Valdez, T. and Horbatsch, M. and Marsman, A. and Vutha, A. C. and Hessels, E. A.",
    title = "{A measurement of the atomic hydrogen Lamb shift and the proton charge radius}",
    doi = "10.1126/science.aau7807",
    journal = "Science",
    volume = "365",
    number = "6457",
    pages = "1007--1012",
    year = "2019"
}

@article{Heacock:2021btd,
    author = "Heacock, Benjamin and others",
    title = {{Pendell{\"o}sung interferometry probes the neutron charge radius, lattice dynamics, and fifth forces}},
    eprint = "2103.05428",
    archivePrefix = "arXiv",
    primaryClass = "nucl-ex",
    doi = "10.1126/science.abc2794",
    journal = "Science",
    volume = "373",
    number = "6560",
    pages = "abc2794",
    year = "2021"
}

@article{Gao:2021sml,
    author = "Gao, Haiyan and Vanderhaeghen, Marc",
    title = "{The proton charge radius}",
    eprint = "2105.00571",
    archivePrefix = "arXiv",
    primaryClass = "hep-ph",
    doi = "10.1103/RevModPhys.94.015002",
    journal = "Rev. Mod. Phys.",
    volume = "94",
    number = "1",
    pages = "015002",
    year = "2022"
}

@article{Atac:2021wqj,
    author = "Atac, H. and Constantinou, M. and Meziani, Z. -E and Paolone, M. and Sparveris, N.",
    title = "{Measurement of the neutron charge radius and the role of its constituents}",
    eprint = "2103.10840",
    archivePrefix = "arXiv",
    primaryClass = "nucl-ex",
    doi = "10.1038/s41467-021-22028-z",
    journal = "Nature Commun.",
    volume = "12",
    number = "1",
    pages = "1759",
    year = "2021",
    note = "[Erratum: Nature Commun. 12, 3290 (2021)]"
}

@article{Xiong:2023zih,
    author = "Xiong, Weizhi and Peng, Chao",
    title = "{Proton Electric Charge Radius from Lepton Scattering}",
    eprint = "2302.13818",
    archivePrefix = "arXiv",
    primaryClass = "nucl-ex",
    doi = "10.3390/universe9040182",
    journal = "Universe",
    volume = "9",
    number = "4",
    pages = "182",
    year = "2023"
}

@article{Lin:2024bzo,
    author = "Lin, Yong-Hui and Guo, Feng-Kun and Mei{\ss}ner, Ulf-G.",
    title = "{The proton charge radius from dimuon photoproduction off the proton}",
    eprint = "2407.20375",
    archivePrefix = "arXiv",
    primaryClass = "hep-ph",
    doi = "10.1016/j.physletb.2024.139023",
    journal = "Phys. Lett. B",
    volume = "858",
    pages = "139023",
    year = "2024"
}

@article{Aliev:2008cs,
    author = "Aliev, T. M. and Azizi, K. and Ozpineci, A. and Savci, M.",
    title = "{Nucleon Electromagnetic Form Factors in QCD}",
    eprint = "0802.3008",
    archivePrefix = "arXiv",
    primaryClass = "hep-ph",
    doi = "10.1103/PhysRevD.77.114014",
    journal = "Phys. Rev. D",
    volume = "77",
    pages = "114014",
    year = "2008"
}

@article{Bauer:2012pv,
    author = "Bauer, T. and Bernauer, J. C. and Scherer, S.",
    title = "{Electromagnetic form factors of the nucleon in effective field theory}",
    eprint = "1209.3872",
    archivePrefix = "arXiv",
    primaryClass = "nucl-th",
    doi = "10.1103/PhysRevC.86.065206",
    journal = "Phys. Rev. C",
    volume = "86",
    pages = "065206",
    year = "2012"
}

@article{Anikin:2013aka,
    author = "Anikin, I. V. and Braun, V. M. and Offen, N.",
    title = "{Nucleon Form Factors and Distribution Amplitudes in QCD}",
    eprint = "1310.1375",
    archivePrefix = "arXiv",
    primaryClass = "hep-ph",
    doi = "10.1103/PhysRevD.88.114021",
    journal = "Phys. Rev. D",
    volume = "88",
    pages = "114021",
    year = "2013"
}

@article{Sharma:2013mfa,
    author = "Sharma, Neetika and Dahiya, Harleen",
    title = "{Charge radii of octet and decuplet baryons in chiral constituent quark model}",
    doi = "10.1007/s12043-013-0575-7",
    journal = "Pramana",
    volume = "81",
    number = "3",
    pages = "449--465",
    year = "2013"
}

@article{Flores-Mendieta:2015wir,
    author = "Flores-Mendieta, Rub{\'e}n and Rivera-Ruiz, Mayra Alejandra",
    title = "{Dirac form factors and electric charge radii of baryons in the combined chiral and 1/N$_c$ expansions}",
    eprint = "1511.02932",
    archivePrefix = "arXiv",
    primaryClass = "hep-ph",
    doi = "10.1103/PhysRevD.92.094026",
    journal = "Phys. Rev. D",
    volume = "92",
    number = "9",
    pages = "094026",
    year = "2015"
}

@article{He:2017viu,
    author = "He, Fangcheng and Wang, P.",
    title = "{Nucleon electromagnetic form factors with a nonlocal chiral effective Lagrangian}",
    eprint = "1711.05896",
    archivePrefix = "arXiv",
    primaryClass = "nucl-th",
    doi = "10.1103/PhysRevD.97.036007",
    journal = "Phys. Rev. D",
    volume = "97",
    number = "3",
    pages = "036007",
    year = "2018"
}

@article{Sufian:2016hwn,
    author = {Sufian, Raza Sabbir and de T{\'e}ramond, Guy F. and Brodsky, Stanley J. and Deur, Alexandre and Dosch, Hans G{\"u}nter},
    title = "{Analysis of nucleon electromagnetic form factors from light-front holographic QCD : The spacelike region}",
    eprint = "1609.06688",
    archivePrefix = "arXiv",
    primaryClass = "hep-ph",
    reportNumber = "JLAB-PHY-16-2363, SLAC--PUB--16806",
    doi = "10.1103/PhysRevD.95.014011",
    journal = "Phys. Rev. D",
    volume = "95",
    number = "1",
    pages = "014011",
    year = "2017"
}

@article{QCDSF:2017ssq,
    author = "Chambers, A. J. and others",
    collaboration = "QCDSF, UKQCD, CSSM",
    title = "{Electromagnetic form factors at large momenta from lattice QCD}",
    eprint = "1702.01513",
    archivePrefix = "arXiv",
    primaryClass = "hep-lat",
    reportNumber = "ADP-17-06-T1012, DESY-17-019, Edinburgh-2017-02, LTH-1119",
    doi = "10.1103/PhysRevD.96.114509",
    journal = "Phys. Rev. D",
    volume = "96",
    number = "11",
    pages = "114509",
    year = "2017"
}

@article{Hasan:2017wwt,
    author = "Hasan, Nesreen and Green, Jeremy and Meinel, Stefan and Engelhardt, Michael and Krieg, Stefan and Negele, John and Pochinsky, Andrew and Syritsyn, Sergey",
    title = "{Computing the nucleon charge and axial radii directly at $Q^2=0$ in lattice QCD}",
    eprint = "1711.11385",
    archivePrefix = "arXiv",
    primaryClass = "hep-lat",
    reportNumber = "RBRC-1255, DESY-17-203, RBRC-1261",
    doi = "10.1103/PhysRevD.97.034504",
    journal = "Phys. Rev. D",
    volume = "97",
    number = "3",
    pages = "034504",
    year = "2018"
}

@article{HillerBlin:2017syu,
    author = "Hiller Blin, A. N.",
    title = "{Systematic study of octet-baryon electromagnetic form factors in covariant chiral perturbation theory}",
    eprint = "1707.02255",
    archivePrefix = "arXiv",
    primaryClass = "hep-ph",
    doi = "10.1103/PhysRevD.96.093008",
    journal = "Phys. Rev. D",
    volume = "96",
    number = "9",
    pages = "093008",
    year = "2017"
}

@article{Alexandrou:2018sjm,
    author = "Alexandrou, C. and Bacchio, S. and Constantinou, M. and Finkenrath, J. and Hadjiyiannakou, K. and Jansen, K. and Koutsou, G. and Vaquero Aviles-Casco, A.",
    title = "{Proton and neutron electromagnetic form factors from lattice QCD}",
    eprint = "1812.10311",
    archivePrefix = "arXiv",
    primaryClass = "hep-lat",
    reportNumber = "DESY-18-033, DESY 18-033",
    doi = "10.1103/PhysRevD.100.014509",
    journal = "Phys. Rev. D",
    volume = "100",
    number = "1",
    pages = "014509",
    year = "2019"
}

@article{Alarcon:2018zbz,
    author = "Alarc{\'o}n, J. M. and Higinbotham, D. W. and Weiss, C. and Ye, Z.",
    title = "{Proton charge radius extraction from electron scattering data using dispersively improved chiral effective field theory}",
    eprint = "1809.06373",
    archivePrefix = "arXiv",
    primaryClass = "hep-ph",
    reportNumber = "JLAB-THY-18-2804",
    doi = "10.1103/PhysRevC.99.044303",
    journal = "Phys. Rev. C",
    volume = "99",
    number = "4",
    pages = "044303",
    year = "2019"
}

@article{Shintani:2018ozy,
    author = "Shintani, Eigo and Ishikawa, Ken-Ichi and Kuramashi, Yoshinobu and Sasaki, Shoichi and Yamazaki, Takeshi",
    title = "{Nucleon form factors and root-mean-square radii on a (10.8  fm)$^4$ lattice at the physical point}",
    eprint = "1811.07292",
    archivePrefix = "arXiv",
    primaryClass = "hep-lat",
    doi = "10.1103/PhysRevD.99.014510",
    journal = "Phys. Rev. D",
    volume = "99",
    number = "1",
    pages = "014510",
    year = "2019",
    note = "[Erratum: Phys.Rev.D 102, 019902 (2020)]"
}

@article{Filin:2019eoe,
    author = {Filin, A. A. and Baru, V. and Epelbaum, E. and Krebs, H. and M{\"o}ller, D. and Reinert, P.},
    title = "{Extraction of the neutron charge radius from a precision calculation of the deuteron structure radius}",
    eprint = "1911.04877",
    archivePrefix = "arXiv",
    primaryClass = "nucl-th",
    doi = "10.1103/PhysRevLett.124.082501",
    journal = "Phys. Rev. Lett.",
    volume = "124",
    number = "8",
    pages = "082501",
    year = "2020"
}

@article{Mondal:2019jdg,
    author = "Mondal, Chandan and Xu, Siqi and Lan, Jiangshan and Zhao, Xingbo and Li, Yang and Chakrabarti, Dipankar and Vary, James P.",
    title = "{Proton structure from a light-front Hamiltonian}",
    eprint = "1911.10913",
    archivePrefix = "arXiv",
    primaryClass = "hep-ph",
    doi = "10.1103/PhysRevD.102.016008",
    journal = "Phys. Rev. D",
    volume = "102",
    number = "1",
    pages = "016008",
    year = "2020"
}

@article{Jang:2019jkn,
    author = "Jang, Yong-Chull and Gupta, Rajan and Lin, Huey-Wen and Yoon, Boram and Bhattacharya, Tanmoy",
    title = "{Nucleon electromagnetic form factors in the continuum limit from ( 2+1+1 )-flavor lattice QCD}",
    eprint = "1906.07217",
    archivePrefix = "arXiv",
    primaryClass = "hep-lat",
    reportNumber = "LA-UR-19-25275, MSUHEP-19-006",
    doi = "10.1103/PhysRevD.101.014507",
    journal = "Phys. Rev. D",
    volume = "101",
    number = "1",
    pages = "014507",
    year = "2020"
}

@article{Barabanov:2020jvn,
    author = "Barabanov, M. Yu. and others",
    title = "{Diquark correlations in hadron physics: Origin, impact and evidence}",
    eprint = "2008.07630",
    archivePrefix = "arXiv",
    primaryClass = "hep-ph",
    reportNumber = "NJU-INP 024/20, JLAB-PHY-21-3316",
    doi = "10.1016/j.ppnp.2020.103835",
    journal = "Prog. Part. Nucl. Phys.",
    volume = "116",
    pages = "103835",
    year = "2021"
}

@article{Yang:2020rpi,
    author = "Yang, Mingyang and Wang, Ping",
    title = "{Electromagnetic form factors of octet baryons with the nonlocal chiral effective theory}",
    eprint = "2005.11971",
    archivePrefix = "arXiv",
    primaryClass = "hep-ph",
    doi = "10.1103/PhysRevD.102.056024",
    journal = "Phys. Rev. D",
    volume = "102",
    number = "5",
    pages = "056024",
    year = "2020"
}

@article{Contreras:2021epz,
    author = "Contreras, Miguel Angel Martin and Capossoli, Eduardo Folco and Li, Danning and Vega, Alfredo and Boschi-Filho, Henrique",
    title = "{Proton and neutron form factors from deformed gravity/gauge duality}",
    eprint = "2108.05427",
    archivePrefix = "arXiv",
    primaryClass = "hep-ph",
    doi = "10.1016/j.physletb.2021.136638",
    journal = "Phys. Lett. B",
    volume = "822",
    pages = "136638",
    year = "2021"
}

@article{Mamo:2021jhj,
    author = "Mamo, Kiminad A. and Zahed, Ismail",
    title = "{Electromagnetic radii of the nucleon in soft-wall holographic QCD}",
    eprint = "2106.00752",
    archivePrefix = "arXiv",
    primaryClass = "hep-ph",
    doi = "10.1016/j.nuclphysb.2023.116388",
    journal = "Nucl. Phys. B",
    volume = "997",
    pages = "116388",
    year = "2023"
}

@article{Bar:2021crj,
    author = "Bar, Oliver and Colic, Haris",
    title = "{N{\ensuremath{\pi}}-state contamination in lattice calculations of the nucleon electromagnetic form factors}",
    eprint = "2104.00329",
    archivePrefix = "arXiv",
    primaryClass = "hep-lat",
    reportNumber = "HU-EP-21/04",
    doi = "10.1103/PhysRevD.103.114514",
    journal = "Phys. Rev. D",
    volume = "103",
    number = "11",
    pages = "114514",
    year = "2021"
}

@article{Brodsky:2022kef,
    author = "Brodsky, Stanley J. and Lyubovitskij, Valery E. and Schmidt, Ivan",
    title = "{Heavy quark contribution to the electromagnetic properties of the nucleon}",
    eprint = "2209.00403",
    archivePrefix = "arXiv",
    primaryClass = "hep-ph",
    reportNumber = "SLAC-PUB-17698",
    doi = "10.1103/PhysRevD.106.094025",
    journal = "Phys. Rev. D",
    volume = "106",
    number = "9",
    pages = "094025",
    year = "2022"
}

@article{Ding:2022ows,
    author = "Ding, Minghui and Roberts, Craig D. and Schmidt, Sebastian M.",
    title = "{Emergence of Hadron Mass and Structure}",
    eprint = "2211.07763",
    archivePrefix = "arXiv",
    primaryClass = "hep-ph",
    reportNumber = "NJU-INP 066/22",
    doi = "10.3390/particles6010004",
    journal = "Particles",
    volume = "6",
    number = "1",
    pages = "57--120",
    year = "2023"
}

@article{Djukanovic:2023beb,
    author = "Djukanovic, Dalibor and von Hippel, Georg and Meyer, Harvey B. and Ottnad, Konstantin and Salg, Miguel and Wittig, Hartmut",
    title = "{Electromagnetic form factors of the nucleon from Nf=2+1 lattice QCD}",
    eprint = "2309.06590",
    archivePrefix = "arXiv",
    primaryClass = "hep-lat",
    reportNumber = "MITP-23-044",
    doi = "10.1103/PhysRevD.109.094510",
    journal = "Phys. Rev. D",
    volume = "109",
    number = "9",
    pages = "094510",
    year = "2024"
}

@article{Chen:2023dxp,
    author = "Chen, Yi and Lorc{\'e}, C{\'e}dric",
    title = "{Nucleon relativistic polarization and magnetization distributions}",
    eprint = "2302.04672",
    archivePrefix = "arXiv",
    primaryClass = "hep-ph",
    doi = "10.1103/PhysRevD.107.096003",
    journal = "Phys. Rev. D",
    volume = "107",
    number = "9",
    pages = "096003",
    year = "2023"
}

@article{Alvarado:2023loi,
    author = "Alvarado, Fernando and An, Di and Alvarez-Ruso, Luis and Leupold, Stefan",
    title = "{Light quark mass dependence of nucleon electromagnetic form factors in dispersively modified chiral perturbation theory}",
    eprint = "2310.07796",
    archivePrefix = "arXiv",
    primaryClass = "hep-ph",
    doi = "10.1103/PhysRevD.108.114021",
    journal = "Phys. Rev. D",
    volume = "108",
    number = "11",
    pages = "114021",
    year = "2023"
}

@article{Tsuji:2023llh,
    author = "Tsuji, Ryutaro and Aoki, Yasumichi and Ishikawa, Ken-Ichi and Kuramashi, Yoshinobu and Sasaki, Shoichi and Sato, Kohei and Shintani, Eigo and Watanabe, Hiromasa and Yamazaki, Takeshi",
    collaboration = "PACS",
    title = "{Nucleon form factors in Nf=2+1 lattice QCD at the physical point: Finite lattice spacing effect on the root-mean-square radii}",
    eprint = "2311.10345",
    archivePrefix = "arXiv",
    primaryClass = "hep-lat",
    reportNumber = "UTHEP-783, UTCCS-P-149, HUPD-2307, YITP-23-143",
    doi = "10.1103/PhysRevD.109.094505",
    journal = "Phys. Rev. D",
    volume = "109",
    number = "9",
    pages = "094505",
    year = "2024"
}

@article{Miller:2007uy,
    author = "Miller, Gerald A.",
    title = "{Charge Density of the Neutron}",
    eprint = "0705.2409",
    archivePrefix = "arXiv",
    primaryClass = "nucl-th",
    reportNumber = "NT@UW-07-07",
    doi = "10.1103/PhysRevLett.99.112001",
    journal = "Phys. Rev. Lett.",
    volume = "99",
    pages = "112001",
    year = "2007"
}

@article{Hill:2010yb,
    author = "Hill, Richard J. and Paz, Gil",
    title = "{Model independent extraction of the proton charge radius from electron scattering}",
    eprint = "1008.4619",
    archivePrefix = "arXiv",
    primaryClass = "hep-ph",
    reportNumber = "EFI-PREPRINT-10-21",
    doi = "10.1103/PhysRevD.82.113005",
    journal = "Phys. Rev. D",
    volume = "82",
    pages = "113005",
    year = "2010"
}

@article{Griffioen:2015hta,
    author = "Griffioen, Keith and Carlson, Carl and Maddox, Sarah",
    title = "{Consistency of electron scattering data with a small proton radius}",
    eprint = "1509.06676",
    archivePrefix = "arXiv",
    primaryClass = "nucl-ex",
    doi = "10.1103/PhysRevC.93.065207",
    journal = "Phys. Rev. C",
    volume = "93",
    number = "6",
    pages = "065207",
    year = "2016"
}

@article{Lorenz:2014vha,
    author = "Lorenz, I. T. and Mei{\ss}ner, Ulf-G.",
    title = "{Reduction of the proton radius discrepancy by 3{\ensuremath{\sigma}}}",
    eprint = "1406.2962",
    archivePrefix = "arXiv",
    primaryClass = "hep-ph",
    doi = "10.1016/j.physletb.2014.08.010",
    journal = "Phys. Lett. B",
    volume = "737",
    pages = "57--59",
    year = "2014"
}

@article{Graczyk:2014lba,
    author = "Graczyk, Krzysztof M. and Juszczak, Cezary",
    title = "{Proton radius from Bayesian inference}",
    eprint = "1408.0150",
    archivePrefix = "arXiv",
    primaryClass = "hep-ph",
    doi = "10.1103/PhysRevC.90.054334",
    journal = "Phys. Rev. C",
    volume = "90",
    pages = "054334",
    year = "2014"
}

@article{Higinbotham:2015rja,
    author = "Higinbotham, Douglas W. and Kabir, Al Amin and Lin, Vincent and Meekins, David and Norum, Blaine and Sawatzky, Brad",
    title = "{Proton radius from electron scattering data}",
    eprint = "1510.01293",
    archivePrefix = "arXiv",
    primaryClass = "nucl-ex",
    reportNumber = "JLAB-PHY-16-2",
    doi = "10.1103/PhysRevC.93.055207",
    journal = "Phys. Rev. C",
    volume = "93",
    number = "5",
    pages = "055207",
    year = "2016"
}

@article{Lee:2015jqa,
    author = "Lee, Gabriel and Arrington, John R. and Hill, Richard J.",
    title = "{Extraction of the proton radius from electron-proton scattering data}",
    eprint = "1505.01489",
    archivePrefix = "arXiv",
    primaryClass = "hep-ph",
    reportNumber = "EFI-PREPRINT-14-35",
    doi = "10.1103/PhysRevD.92.013013",
    journal = "Phys. Rev. D",
    volume = "92",
    number = "1",
    pages = "013013",
    year = "2015"
}

@article{Horbatsch:2016ilr,
    author = "Horbatsch, Marko and Hessels, Eric A. and Pineda, Antonio",
    title = "{Proton radius from electron-proton scattering and chiral perturbation theory}",
    eprint = "1610.09760",
    archivePrefix = "arXiv",
    primaryClass = "nucl-th",
    doi = "10.1103/PhysRevC.95.035203",
    journal = "Phys. Rev. C",
    volume = "95",
    number = "3",
    pages = "035203",
    year = "2017"
}

@article{Sick:2017aor,
    author = "Sick, Ingo and Trautmann, Dirk",
    title = "{Proton rms-radii from low-q power expansions?}",
    eprint = "1701.01809",
    archivePrefix = "arXiv",
    primaryClass = "nucl-ex",
    doi = "10.1103/PhysRevC.95.012501",
    journal = "Phys. Rev. C",
    volume = "95",
    number = "1",
    pages = "012501",
    year = "2017"
}

\end{document}